\documentclass[openacc]{rstransa}
\usepackage{tom}
\usepackage{amsmath}
\usepackage{amssymb}
\usepackage{graphicx}
\author{Thomas Engl, Juan Diego Urbina and Klaus Richter}
 \title{The semiclassical propagator in Fock space: \\
dynamical echo and many-body interference}
\address{Institut f\"ur Theoretische Physik, Universit\"at Regensburg, D-93040 Regensburg, Germany}


\corres{Thomas Engl\\
\email{thomas.engl@physik.uni-regensburg.de}}

\begin{document}
\begin{abstract}
We present a semiclassical approach to many-body quantum propagation in terms of coherent sums over quantum amplitudes associated with the solutions of corresponding classical nonlinear wave equations. This approach adequately describes interference effects in the many-body space of interacting bosonic systems.

The main quantity of interest, the transition amplitude between Fock states when the dynamics is driven by both single-particlecontributions and many-body interactions of similar magnitude, is non-perturbatively constructed in the spirit of Gutzwiller's derivation of the van Vleck propagator from the path integral representation of the time evolution operator, but lifted to the space of symmetrized many-body states. Effects beyond mean-field, here representing the classical limit of the theory, are  semiclassically described by means of interfering amplitudes where the action and stability of the classical solutions enter. In this way, a genuinely many-body echo phenomenon, coherent backscattering in Fock space, is presented arising due to coherent quantum interference between classical solutions related by time reversal.
\end{abstract}
\maketitle
%
%
%
%
%
%
%%%%%%%%%% Insert the texts which can accomdate on firstpage in the tag "fmtext" %%%%%
%
%\begin{fmtext}
\section{Introduction}
Interference effects, as a hallmark of quantum behavior, are so essential for the description of the fundamental physical processes around us that the founding fathers of quantum mechanics included interference already at the kinematical level of modern quantum theory in the form of a central postulate: the state space structure of quantum mechanics is that of a vector space, and therefore the sum of two possible states represents also a valid state. This kinematic concept extends its implications into the domain of unitary dynamics in isolated quantum systems, which is postulated to be represented by linear operators acting on state vectors and therefore allowing for the coherent superposition of different histories when describing a single physical process. Given the fragile nature of quantum coherent effects, much has been learned from the formal correspondence between classical field theory, in particular from Maxwel's electrodynamics, and the solutions of both the stationary and time-dependent 
single-particle Schr\"odinger equation describing (predominantly) the orbital degrees of freedom of a particle in real space. Many of the benchmark coherent effects, now routinely studied in the framework of quantum transport, had been, for instance, first observed within a purely classical wave context. In this respect, echo phenomena do not make an exception: there is a perfect correspondence between classical wave echoes and quantum wave mechanical signatures of coherent superpositions that reveal the information about the initial state.

In exact analogy with classical wave systems where ray dynamics serves as a backbone supporting wave propagation in the limit of short wavelengths, wave mechanics of single-particle quantum systems can be described in the limit 
of large classical actions (compared to $\hbar$) by an asymptotic version of quantum mechanics where only classical information is required (apart from $\hbar$). This semiclassical program, initiated already by Bohr and cast in its elaborate, modern form by Gutzwiller\cite{Gutzwiller}, accomplishes this idea to its ultimate limits: not only the average incoherent aspects of a quantum process are given by their classical expectations (in accordance with the correspondence principle) but also the quantum fluctuations around them can be unambiguously calculated using only classical information. The seminal contributions of van Vleck, Feynman and Gutzwiller consist in the explicit construction of the machinery transforming classical information into quantum mechanical predictions. The ultimate success of the theory, based in particular on the underlying propagator, is reflected in the physical understanding and quantitative analysis of quantum interference phenomena in terms of classical solutions, their actions and their stabilities.

The fruitful analogy between quantum amplitudes and classical fields cannot be simply lifted to the many-body level where effective single-particle description ceases to be valid. For interacting many-body systems the quantum state is by definition a high-dimensional object: many-body interference takes place in the space of quantum states without a real space analogue. Therefore the obvious question arises whether it is possible to export the highly successful semiclassical techniques into the realm of interacting quantum systems. More precise questions concern the very existence of a semiclassical limit and the technical issue of the implementation of a Gutzwiller approach in many-body space.

This sort of questions might have remained to be of rather academic nature, if the experimental studies of interference effects, where semiclassics has proven its value, would not advance so rapidly. The successful realization of genuine many-body coherent effects like Rabi oscillations and Anderson localization in cold atom quantum gases, and first steps towards Fock space tomography of cold atoms in optical lattices \cite{Greiner, Bloch} further trigger theoretical methods able to capture many-body interference phenomena in its characteristic non-perturbative form.
 
Since the classical limit of systems of interacting bosonic cold atoms, representing discrete bosonic quantum fields, has been known for a while in the form of a non-linear classical field equation, one might think that a
semiclassical approach could have been easily implemented. However, in the context of interacting bosons this Gross-Pitaevskii equation is derived from a variational approach describing the ground state of the system \cite{MF1} and, although within this regime it has enjoyed remarkable success, it has been implicitly assumed that it cannot by construction deal with full quantum effects and that it is restricted to ground-state properties. Here, we show that the same mean-field equation emerges as the rigorous limit of a quantum theory of interacting bosons when the latter is analyzed semiclassically within a field-theoretical formalism. In this approach, one finds naturally that the classical limit corresponds to $N\to\infty$, where $N$ is the total number of particles, thus providing the connection to the standard derivations of the mean-field equations. Lifting the Gross-Pitaevskii equation from a mean field picture to the notion of a classical limit makes it amenable to semiclassical appraoches and thereby allows for using classical information encoded in the mean field equations 
in an alternative way to describe both many-body interference and quantum effects of excited states beyond the mean field description.
 
Here we provide a detailed account of such a semiclassical theory of interacting bosonic fields, where quantum indistinguishability and inter-particle interactions are accounted for. This approach that has been acheived only recently \cite{cbs_fock} follows and extends Gutzwiller's work \cite{Gutzwiller} based on the asymptotic analysis of the exact path integral representation of the quantum propagator within a suitable basis. Equipped  with a van Vleck-Gutzwiller-type propagator in Fock space, we will apply this approach to the calculation of an echo effect due to the coherent interference between quantum amplitudes associated with different classical mean-field solutions related by time-reversal symmetry. We dubbed this novel many-body echo effect {\em Coherent Backscattering (CBS) in Fock space}.

Echoes serve as prominent probes for unitary quantum evolution and in particular deviations from it, with spin \cite{Hahn-echo} and polarization \cite{Zhang1992} echoes in NMR as ground breaking examples. The generalization of these concepts, originally devoted to spin-like properties, to the quantum time evolution and in particular the study of its sensitivity with respect to perturbations has led to the closely related notions of fidelity~\cite{loschmidt_echo_introduction} and Loschmidt echo~\cite{Usaj} and a whole corresponding respective research field (for reviews see~\cite{Gorin2006, Jacquod2009, scholarpedia_loschmidt}). This broad interest in echo phenomena was partly triggered by the pioneering work of Jalabert and Pastawski~\cite{loschmidt_rodolfo} who considered the Loschmidt echo related to (the overlap of) the spatial degrees of freedom of wave functions in wave-chaotic single-particle systems employing a semiclassical approximation of the propagator. Here we lift such a semiclassical method to the level of propagation in many-body Fock space; however, we do not consider an echo upon time inversion as in the case above, but instead a many-body generalization of dynamical echoes~\cite{Prigodin1994,Pastawski1995} such as coherent backscattering. 

To this end we provide a comprehensive and detailed derivation of the semiclassical propagator in Fock space that
was only breifly outlined in Ref.~\cite{cbs_fock}. The paper is structured as follows: In Sect.~\ref{sec:propagator_quadrature} we introduce so called quadrature eigenstates, which are used to first derive a path integral and and then a semiclassical approximation for the propagator. After that, we demonstrate how to perform a basis change from quadrature to Fock states in Sect.~\ref{sec:basis-change}. Finally, the resulting semiclassical propagator is used to derive coherent backscattering in Fock space in Sect.~\ref{sec:cbs}, before we conclude in Sect.~\ref{sec:conclusion}
%
%
%\end{fmtext}
%
%
%\maketitle
%
%
%
%
%
\section{The propagator in quadrature representation}
\label{sec:propagator_quadrature}
\subsection{Quadrature states}
In this section, we will introduce the quadrature eigenstates and their most important properties without any proof. For more details we refer the reader \eg to \cite{VogelWelsch200607}.

The quadrature states $\ket{\V{q}}$ and $\ket{\V{p}}$ are defined as the eigenstates of the quadrature operator,
\begin{subequations}
\begin{eqnarray}
&\hat{q}_{\sites}\ket{\V{q}}=q_{\sites}\ket{\V{q}}, & \hat{p}_{\sites}\ket{\V{p}}=p_{\sites}\ket{\V{p}}, \\
&\hat{q}_{\sites}=\quadfac\rbr{\hat{a}_{\sites}^{}+\hat{a}_{\sites}^\dagger}, &\hat{p}_{\sites}=-\rmi\quadfac\rbr{\hat{a}_{\sites}^{}-\hat{a}_{\sites}^\dagger},
\end{eqnarray}
\end{subequations}
where $\quadfac$ is for now an arbitrary, but real parameter and $\hat{a}_{\sites}^\dagger$ and $\hat{a}_{\sites}^{}$ are the ${\sites}$th creation and annihilation operators, respectively.

%The so defined quadrature operators $\hat{q}_{\sites},\hat{p}_{\sites}$ satisfy the commutation relations
%%
%
%\begin{equation}
%\cbr{\hat{q}_{\sites},\hat{q}_{\sites^{\prime}}}=\cbr{\hat{p}_{\sites},\hat{p}_{\sites^{\prime}}}=0, \qquad \cbr{\hat{q}_{\sites},\hat{p}_{\sites^{\prime}}}=2\rmi\quadfac^2\delta_{\sites\sites^{\prime}}.
%\label{eq:commutation_quadrature}
%\end{equation}
%
The quadrature eigenstates can be expanded in Fock states $\ket{\V{n}}$, defined as usual as the normalized solutions of the eigenvalue equations $\hat{n}_{\sites}\ket{\V{n}}=\hat{a}_{\sites}^\dagger\hat{a}_{\sites}^{}\ket{\V{n}}=n_{\sites}\ket{\V{n}}$, by using their overlaps
\begin{subequations}
\begin{align}
\olap{\V{n}}{\V{q}}&=\prodl{\sites}{}\frac{\exp\rbr{-\frac{q_{\sites}^2}{4\quadfac^2}}}{\sqrt{2^{n_{\sites}}n_{\sites}!\sqrt{2\pi}\quadfac}}H_{n_{\sites}}\rbr{\frac{q_{\sites}}{\sqrt{2}\quadfac}}, \\
\olap{\V{n}}{\V{p}}&=\prodl{\sites}{}\frac{\exp\rbr{-\frac{p_{\sites}^2}{4\quadfac^2}}\rmi^{n_{\sites}}}{\sqrt{2^{n_{\sites}}n_{\sites}!\sqrt{2\pi}\quadfac}}H_{n_{\sites}}\rbr{\frac{p_{\sites}}{\sqrt{2}\quadfac}},
\end{align}
\end{subequations}
where $H_n$ denotes the $n$-th Hermite polynomial. For large occupation numbers, $n_{\sites}\gg1$, these overlaps can in the oscillatory region, $\abs{q_l}\leq2\quadfac\sqrt{n_l+1/2}$, be approximated by the WKB-approximation \cite{Brack+Bhaduri},
\begin{equation}
\olap{\V{n}}{\V{q}}\approx\prodl{\sites}{}\sqrt{\frac{2}{\pi\sqrt{4\quadfac^2\rbr{n_{\sites}+\frac{1}{2}}-q_{\sites}^2}}}\cos\cbr{F\rbr{q_{\sites},n_{\sites}}+\frac{\pi}{4}},
\label{eq:assymptotics_overlap}
\end{equation}
where $F(q,n)$ is the generating function for the canonical transformation from $(q,p)$ to $(n,\theta)$ with $q=2\quadfac\sqrt{n+\frac{1}{2}}\cos\theta$ and $p=2\quadfac\sqrt{n+\frac{1}{2}}\sin\theta$ and is given by
\begin{equation}
F\rbr{q,n}=\frac{q}{4\quadfac^2}\sqrt{4\quadfac^2\rbr{n+\frac{1}{2}}-q^2}-\rbr{n+\frac{1}{2}}\arccos\rbr{\frac{q}{2\quadfac\sqrt{n+\frac{1}{2}}}}.
\end{equation}
It is worth to notice that the generating function can also be written as an integral along a classical path,
\begin{equation}
F\rbr{q,n}=\frac{1}{\quadfac}\intg{q_0}{q}{x}\sqrt{n+\frac{1}{2}-\frac{x^2}{4\quadfac^2}},
\end{equation}
where $q_0=2\quadfac\sqrt{n+1/2}$.
As a generating function of a canonical transformation, the derivatives of $F(q,n)$ satisfy
\begin{subequations}
\begin{align}
\pdiff{F(q,n)}{q}&=\frac{1}{\quadfac}\sqrt{n+\frac{1}{2}-\frac{q^2}{4\quadfac^2}}=\frac{\abs{p}}{2\quadfac^2}
\label{eq:deriv_generating_q-n_q} \\
\pdiff{F(q,n)}{n}&=-\arccos\rbr{\frac{q}{2\quadfac\sqrt{n+\frac{1}{2}}}}=-\theta
\label{eq:deriv_generating_q-n_n}
\end{align}
\end{subequations}
Note that, by this definition of $\theta$, the phase is only defined between 0 and $\pi$, and therefore $\abs{p}=p$.

The overlap of different quadrature eigenstates are given by
\begin{subequations}
\label{eq:overlaps_quadratures}
\begin{align}
\olap{\V{q}}{\V{q}^\prime}&=\delta\rbr{\V{q}-\V{q}^\prime},
\label{eq:olap_q-q} \\
\olap{\V{p}}{\V{p}^\prime}&=\delta\rbr{\V{p}-\V{p}^\prime},
\label{eq:olap_p-p} \\
\olap{\V{q}}{\V{p}}&=\prodl{l}{}\frac{1}{\sqrt{4\quadfac^2\pi}}\exp\rbr{\frac{\rmi}{2\quadfac^2}q_lp_l}.
\label{eq:olap_q-p}
\end{align}
\end{subequations}
Finally the resolution of unity in terms of quadrature eigenstates is given by
\begin{equation}
\hat{1}=\int\rmd\V{q}\ket{\V{q}}\bra{\V{q}}=\int\rmd\V{p}\ket{\V{p}}\bra{\V{p}}.
\label{eq:resolution_unity_quadratures}
\end{equation}
Thus, the quadrature states are a proper choice for the construction of a path integral, since the resolution of unity is given by an \emph{integral} rather than a sum. Moreover, for the semiclassical approximation following this construction, they seem to be a better choice than coherent states, since their eigenvalues are real and therefore the complexification necessary for the coherent state propagator can be avoided.
\subsection{The path integral}
In quadrature representation, the exact quantum mechanical propagator is given by the matrix element
\begin{equation}
K\rbr{\qkv{f},\qkv{i};\tf,\ti}=\bra{\qkv{f}}\hat{K}\rbr{\tf,\ti}\ket{\qkv{i}},
\label{eq:def_propagator_quadrature}
\end{equation}
with
\begin{equation}
\hat{K}\rbr{\tf,\ti}=\timeorder\exp\rbr{-\frac{\rmi}{\hbar}\intg{\ti}{\tf}{t}\hat{H}(t)}
\label{eq:quantum_propagator}
\end{equation}
being the quantum mechanical evolution operator for a many-body Hamiltonian $\hat{H}$, which may be \eg a typical Bose-Hubbard Hamiltonian of the form
\begin{equation}
\hat{H}=\sul{\sites,\sites^\prime}{}h_{\sites\sites^\prime}\hat{a}_{\sites}^\dagger\hat{a}_{\sites}^{}+\frac{1}{2}\sul{\sites,\sites^{\prime},\sites^{\prime\prime},\sites^{\prime\prime\prime}}{}U_{\sites\sites^{\prime}\sites^{\prime\prime}\sites^{\prime\prime\prime}}\hat{a}_{\sites}^\dagger\hat{a}_{\sites^\prime}^\dagger\hat{a}_{\sites^{\prime\prime}}^{}\hat{a}_{\sites^{\prime\prime\prime}}^{}.
\label{eq:general_BH}
\end{equation}
The path integral is then constructed by first using Trotter's formula \cite{trotter},
\begin{equation}
\hat{K}\rbr{\tf,\ti}=\pilimit\prodl{\picount=1}{\pisteps}\exp\cbr{-\frac{\rmi\timestep}{\hbar}\hat{H}\rbr{\tf-\picount\timestep}},
\end{equation}
where $\timestep=\rbr{\tf-\ti}/\pisteps$ and inserting one unit operator in terms of quadratures $q$ and $p$, \Eref{eq:resolution_unity_quadratures}, for each time step $\tau$. Following this procedure, however, as already noted in the first quantized case for charged particles in a magnetic field by Feynman in his original derivation of the path integral \cite{Feynman} and reviewed in detail in \cite{pathintegral_magnetic_field}, one has to be careful with the terms of the Hamiltonian containing products of $\hat{q}$ and $\hat{p}$ operators when letting them act on a $\bra{\V{q}}$ or $\bra{\V{p}}$ state, respectively (see also \aref{app:path-integral}). After performing all these steps, in the continuous time limit $\pisteps\to\infty$ the path integral for the propagator in quadrature representation is given by \cite{Schulman,Kleinert}
\begin{equation}
\begin{split}
%K&\rbr{\qkv{f},\qkv{i};\tf,\ti}= \\
%&\pilimit\frac{1}{\rbr{4\pi\quadfac^2}^{\pisteps L}}\intg{}{}{^L\qkj{1}{}}\cdots\intg{}{}{^L\qkj{\pisteps-1}{}}\intg{}{}{^L\pkj{1}{}}\cdots\intg{}{}{^L\pkj{\pisteps}{}} \\
%&\exp\tbr{\rmi\sul{\picount=1}{\pisteps}\cbr{\frac{\pkv{\picount}}{2\quadfac^2}\cdot\rbr{\qkv{\picount}-\qkv{\picount-1}}-\frac{\timestep}{\hbar}\Hcl\rbr{\psikvd{\picount},\psikv{\picount};\ti+\picount\timestep}}},
K&\rbr{\qkv{f},\qkv{i};\tf,\ti}= \\
&\int\limits_{\V{q}\rbr{\ti}=\qkv{i}}^{\V{q}\rbr{\tf}=\qkv{f}}{\rm D}\cbr{\V{q}\rbr{t},\V{p}\rbr{t}}\exp\tbr{\frac{\rm i}{\hbar}\intg{\ti}{\tf}{t}\cbr{\frac{\hbar}{2\quadfac^2}\V{p}\rbr{t}\cdot\dot{\V{q}}\rbr{t}-\Hcl\rbr{\gVd{\psi}\rbr{t},\gV{\psi}\rbr{t};t}}},
\end{split}
\label{eq:many-body_path-integral}
\end{equation}
with %$\psikv{\picount}=\rbr{\qkv{\picount-1}+\rmi\pkv{\picount}}/(2\quadfac)$
$\gV{\psi}\rbr{t}=\rbr{\V{q}\rbr{t}+\rmi\V{p}\rbr{t}}/\rbr{2\quadfac}$ and the boundary conditions %$\qkv{0}=\qkv{i}$ as well as $\qkv{\pisteps}=\qkv{f}$
$\V{q}\rbr{\ti}=\qkv{i}$ as well as $\V{q}\rbr{\tf}=\qkv{f}$. Here, the classical Hamiltonian is defined by
\begin{equation}
\Hcl\rbr{\psivd,\psiv;t}=\frac{\Braket{\V{p}|\hat{H}|\V{q}}}{\Braket{\V{p}|\V{q}}},
\label{eq:classical_hamiltonian_bosons}
\end{equation}
which for a Bose-Hubbard Hamiltonian of the form \eref{eq:general_BH} can be evaluated as
\begin{equation}
\begin{split}
\Hcl&\rbr{\gVd{\psi},\gV{\psi}}= \\
&\sul{\sites,\sites^{\prime}}{}H^{(0)}_{\sites\sites^{\prime}}\rbr{\conjg{\psi}_{\sites}\psi_{\sites^{\prime}}-\frac{1}{2}\delta_{\sites\sites^{\prime}}}+\frac{1}{2}\sul{\sites,\sites^{\prime},\sites^{\prime\prime}\sites^{\prime\prime\prime}}{}U_{\sites\sites^{\prime\prime}\sites^{\prime}\sites^{\prime\prime\prime}}\rbr{\conjg{\psi}_{\sites}\psi_{\sites^{\prime}}-\frac{1}{2}\delta_{\sites\sites^{\prime}}}\rbr{\conjg{\psi}_{\sites^{\prime\prime}}\psi_{\sites^{\prime\prime\prime}}-\frac{1}{2}\delta_{\sites^{\prime\prime}\sites^{\prime\prime\prime}}},
\end{split}
\label{eq:classical_limit_BH}
\end{equation}
which corresponds to the replacement rule
\begin{equation}
\hat{a}_{\sites}^\dagger\hat{a}_{\sites^\prime}^{}\to\conjg{\psi}_{\sites}\psi_{\sites^\prime}^{}-\frac{1}{2}\delta_{\sites\sites^\prime}.
\end{equation}
Note that the classical Hamiltonian also contains constant terms, which give rise to a constant phase in the propagator. Such phases have been found for the semiclassical coherent state propagator \cite{cs-propagator,cs-propagator_severaldim,Garg,Garg+Braun,Pletyukhov_extra-phase,Kochetov,Solari,Vieira+Sacramento}, and have been denoted as Solari-Kochetov extra-phase. It should also be noted that the classical Hamiltonian depends on the chosen ordering of the propagator.
\subsection{Semiclassical approximation}
The semiclassical approximation to the quantum propagator in quadrature representation is obtained by applying the stationay phase approximation to \Eref{eq:many-body_path-integral}. This is to expand the exponent around the stationary points up to second order in $\qkv{\picount}$ and $\pkv{\picount}$ and then perform the thus resulting gaussian integral.

The stationarity conditions in $\qkv{\picount}$ and $\pkv{\picount}$ is equivalent to those in $\psikv{\picount}$ and $\psikvd{\picount}$. In the time-continuous limit $\pisteps\to\infty$, these are given by the equations of motion
\begin{equation}
\rmi\hbar\pdiff{\psiv(t)}{t}=\pdiff{\Hcl\rbr{\psivd(t),\psiv(t);t}}{\psivd(t)}.
\label{eq:eom}
\end{equation}
Thus, the semiclassical propagator in quadrature representation will be given by a sum over \emph{real} mean-field trajectories $\gamma$ obtained by solving the shooting problem consisting of \Eref{eq:eom} under the boundary conditions on the real parts of $\gV{\psi}(t)$,
\begin{equation}
\begin{split}
\Re\psiv\rbr{\ti}&=\frac{1}{2\quadfac}\qkv{i}, \\%\label{eq:bci} \\
\Re\psiv\rbr{\tf}&=\frac{1}{2\quadfac}\qkv{f}. %\label{eq:bcf}
\end{split}
\label{eq:bc_quadratures}
\end{equation}
Evaluating the exponent along the stationary points will yield the phase, the trajectory contributes to the propagator and is given by the classical action of $\gamma$,
\begin{equation}
R_\gamma\rbr{\qkv{f},\qkv{i};\tf,\ti}=\intg{\ti}{\tf}{t}\cbr{\frac{\hbar}{2\quadfac^2}\pv(t)\cdot\dot{\qv}(t)-\Hcl\rbr{\psivd(t),\psiv(t);t}},
\label{eq:action}
\end{equation}
where $\psiv(t)=\rbr{\qv(t)+\rmi\pv(t)}/(2\quadfac)$ is determined by the solution of the equations of motion, \eref{eq:eom}, under the boundary conditions \eref{eq:bc_quadratures}, such that the semiclassical propagator is given by
\begin{equation}
K\rbr{\qkv{f},\qkv{i};\tf,\ti}=\sul{\gamma:\qkv{i}\to\qkv{f}}{}\mathcal{A}_{\gamma}\exp\cbr{\frac{\rmi}{\hbar}R_\gamma\rbr{\qkv{f},\qkv{i};\tf,\ti}}.
\end{equation}
The semiclassical amplitude $\mathcal{A}_{\gamma}$ finally is determined by the integration over quantum fluctuations in the path integral around the stationary (classical) trajectories.
%
%\begin{equation}
%\begin{split}
%\mathcal{A}_{\gamma}=&\pilimit\frac{1}{\rbr{4\pi\quadfac^2}^{\pisteps L}}\intg{}{}{^L\xkj{1}{}}\cdots\intg{}{}{^L\xkj{\pisteps-1}{}}\intg{}{}{^L\ykj{1}{}}\cdots\intg{}{}{^L\ykj{\pisteps}{}} \\
%&\exp\Bigg\{\frac{\rmi}{2\quadfac^2}\ykv{1}\cdot\xkv{1}+\frac{\rmi}{2\quadfac^2}\sul{\picount=2}{\pisteps-1}\ykv{\picount}\cdot\rbr{\xkv{\picount}-\xkv{\picount-1}}-\frac{\rmi}{2\quadfac^2}\ykv{\pisteps}\cdot\xkv{\pisteps-1} \\
%&\qquad-\frac{\rmi\timestep}{\hbar}\sul{\picount=2}{\pisteps}\left(\begin{array}{c} \xkv{\picount-1} \\ \ykv{\picount} \end{array}\right)\pdiff{^2\Hcl\rbr{\psikvd{\picount},\psikv{\picount};\ti+\picount\timestep}}{\left(\xkv{\picount-1},\ykv{\picount}\right)^2}\left(\begin{array}{c} \xkv{\picount-1} \\ \ykv{\picount} \end{array}\right) \\
%&\qquad-\frac{\rmi\timestep}{\hbar}\ykv{1}\pdiff{^2\Hcl\rbr{\psikvd{1},\psikv{1};\ti+\timestep}}{{\ykv{1}}^2}\ykv{1}\Bigg\},
%\end{split}
%\label{eq:semiclassical_prefactor_quad}
%\end{equation}
%
The evaluation of these integrals is very similar to those presented for the semiclassical coherent state propagator in \cite{cs-propagator,cs-propagator_severaldim} and yields (see \aref{app:scl_prefactor_quad})
\begin{equation}
\mathcal{A}_{\gamma}=\sqrt{\det\frac{1}{\rbr{-2\pi\rmi\hbar}}\pdiff{^2R_\gamma\rbr{\qkv{f},\qkv{i};\tf,\ti}}{\qkv{f}\partial\qkv{i}}},
\label{eq:semiclassical_prefactor_quad}
\end{equation}
and thus the semiclassical propagator in quadrature representation is given by
\begin{equation}
\begin{split}
\pushleft{K\rbr{\qkv{f},\qkv{i};\tf,\ti}=} \\
\qquad\sul{\gamma:\qkv{i}\to\qkv{f}}{}\sqrt{\det\frac{1}{\rbr{-2\pi\rmi\hbar}}\pdiff{^2R_\gamma\rbr{\qkv{f},\qkv{i};\tf,\ti}}{\qkv{f}\partial\qkv{i}}}\exp\cbr{\frac{\rmi}{\hbar}R_\gamma\rbr{\qkv{f},\qkv{i};\tf,\ti}}&.
\end{split}
\label{eq:semiclassical_propagator_quadrature}
\end{equation}
%
%\begin{figure}
%\centering
%\includegraphics[width=0.65\textwidth]{bosonic_propagator/figs/classical_trajectory.eps}
%\put(-298,12){$\Re\gV{\psi}=\frac{1}{2\quadfac}\qkv{i}$}
%\put(0,6){$\Re\gV{\psi}=\frac{1}{2\quadfac}\qkv{f}$}
%\put(-135,12){$\rmi\hbar\dot{\gV{\psi}}=\pdiff{\Hcl}{\gVd{\psi}}$}
%\caption{\label{fig:classical_trajectory_quadrature}Graphical representation of a trajectory involved in the semiclassical propagator in quadrature representation.}
%\end{figure}
%
It is important to notice, that here $\qv(t)$ and $\pv(t)$ for the classical trajectories -- in contrast to the coherent state propagators derived in \cite{Garg,cs-propagator,cs-propagator_severaldim,orbital_spin_coherent_states,Garg+Braun} -- are still real variables, since only the initial and final \emph{real parts} are fixed by the boundary conditions, \ie in the approach presented here, no complexification is necessary.

Finally, we would like to state the derivatives of the action with respect to the boundary conditions, which can be calculated by using the equations of motion and are given by
%
%\begin{subnumberedalign}{eq:derivatives_action_bosons}
\begin{equation}
\begin{split}
\pdiff{R_{\gamma}\rbr{\qkv{f},\qkv{i};\tf,\ti}}{\qkj{i}{\sites}}=-\frac{\hbar}{2\quadfac^2}p_\sites^{(\gamma)}\rbr{\ti}, \\ %\label{eq:deriv_qi} \\
\pdiff{R_{\gamma}\rbr{\qkv{f},\qkv{i};\tf,\ti}}{\qkj{f}{\sites}}=\frac{\hbar}{2\quadfac^2}p_\sites^{(\gamma)}\rbr{\tf}. %\label{eq:deriv_qf}
\end{split}
\label{eq:derivatives_action_bosons}
\end{equation}
%\end{subnumberedalign}
%
Furthermore for a time-independent hamiltonian, the time derivative of the action is given by
\begin{equation}
\pdiff{R_{\gamma}\rbr{\qkv{f},\qkv{i};\tf,\ti}}{\tf}=E_\gamma,
\end{equation}
where $E_\gamma=\Hcl\rbr{\psivd(\ti),\psiv(\ti)}$ is the conserved energy of the trajectory.
\section{Basis change to Fock states}
\label{sec:basis-change}
Starting from the semiclassical propagator in quadrature representation, we can now also derive a semiclassical expression for the propagator in Fock representation according to
\begin{equation}
K\rbr{\nfv,\niv;\tf,\ti}=\int\rmd^\Sites\qkv{i}\int\rmd^\Sites\qkv{f}\olap{\nfv}{\qkv{f}}K\rbr{\qkv{f},\qkv{i};\tf,\ti}\olap{\qkv{i}}{\niv}
\label{eq:basis-change_q-n_start}
\end{equation}
by using the asymptotic form \Eref{eq:assymptotics_overlap} for the overlap of a quadrature with a Fock state, neglecting the contributions to the integral from $\abs{\qkj{i/f}{\sites}}>2\quadfac\sqrt{\nifj{\sites}+1/2}$\footnote{The overlap $\Braket{\V{n}|\qv}$ decays exponentially outside this region} and evaluating the integrals in stationary phase approximation.

Using Eqns.~\eref{eq:deriv_generating_q-n_q} and \eref{eq:derivatives_action_bosons}, it is straight forward to find that the stationarity conditions are given by
\begin{subequations}
\label{eq:bc_fock}
\begin{align}
\abs{\psi_{\sites}\rbr{\ti}}^2=&\frac{q_{\sites}^2\rbr{\ti}+p_{\sites}^2\rbr{\ti}}{4\quadfac^2}=\rbr{\nij{\sites}+\frac{1}{2}}, \\
\abs{\psi_{\sites}\rbr{\tf}}^2=&\frac{q_{\sites}^2\rbr{\tf}+p_{\sites}^2\rbr{\tf}}{4\quadfac^2}=\rbr{\nfj{\sites}+\frac{1}{2}}
\end{align}
\end{subequations}
Note that the two possible signs of $\pv\rbr{\ti}$ and $\pv\rbr{\tf}$, which are not encoded in the generating function itself are obtained from the two possible signs of the generating function when expressing the cosine in \Eref{eq:assymptotics_overlap} by exponentials.

However, it turns out, that the boundary conditions \eref{eq:bc_fock} give rise to continuous families of trajectories originating from the $U(1)$-gauge invariance, \ie from the fact, that if $\psiv\rbr{t}$ solves the equations of motion under the boundary conditions \eref{eq:bc_fock}, then $\psiv\rbr{t}\rme^{\rmi\theta}$ also does. Therefore, one of $2\Sites$ the integrations, \eg the one over $\qkj{i}{1}$, cannot be performed in stationary phase approximation.

In fact it turns out to be beneficial to transform the integrations over $\qkj{i/f}{\sites}$ into integrations over $\thetakj{i/f}{\sites}$ defined by
\begin{subequations}
\begin{align}
\qkj{f}{\sites}&=2\quadfac\sqrt{\nfj{\sites}+1/2}\cos\rbr{\thetakj{f}{\sites}+\thetakj{i}{1}}, \\
\qkj{i}{\sites}&=2\quadfac\sqrt{\nij{\sites}+1/2}\cos\rbr{\thetakj{i}{\sites}+\thetakj{i}{1}} \qquad \text{for }\sites>1, \\
\qkj{i}{1}&=2\quadfac\sqrt{\nij{1}+1/2}\cos\rbr{\thetakj{i}{1}}.
\end{align}
\end{subequations}
Then one can perform all the integrations except the one over $\thetakj{i}{1}$ in stationary phase approximation yielding again the stationarity conditions \eref{eq:bc_fock} with
\begin{equation}
\abs{\psi_{1}\rbr{\ti}}^2=\frac{q_{1}^2\rbr{\ti}+p_{1}^2\rbr{\ti}}{4\quadfac^2}=\rbr{\nij{1}+\frac{1}{2}}, \\
\end{equation}
being automatically fulfilled due to conservation of particles. Thus, these integrations select trajectories $\gamma$ satisfying these boundary conditions and additionaly
\begin{equation}
\psi_{1}\rbr{\ti}=\sqrt{\nij{1}+\frac{1}{2}}\rme^{\rmi\thetakj{i}{1}}.
\end{equation}

The evaluation of the exponent at the stationary points yields the classical action of the resulting classical trajectories,
\begin{equation}
R_\gamma\rbr{\nfv,\niv;\ti,\tf}=\intg{\ti}{\tf}{t}\cbr{\hbar\gV{\theta}(t)\cdot\dot{\V{n}}(t)-\Hcl\rbr{\psivd(t),\psiv(t);t}},
\label{eq:action_fock}
\end{equation}
where $\theta_{\sites}(t)$ and $n_{\sites}(t)$ are the real variables defined by
\begin{equation}
\psi_{\sites}(t)=\sqrt{n_{\sites}(t)+\frac{1}{2}}\rme^{\rmi\rbr{\theta_{\sites}(t)+\thetakj{i}{1}}}.
\end{equation}
It is easy to see that -- again due to conservation of particles -- the classical action \eref{eq:action_fock} is independent of the initial global phase $\thetakj{i}{1}$.

Furthermore, the semiclassical prefactor turns out to be independent of $\thetakj{i}{1}$, too, and can be written as
\begin{equation}
\frac{1}{2\pi}\sqrt{{\det}^\prime\frac{1}{\rbr{-2\pi\rmi\hbar}}\pdiff{^2R_{\gamma}\rbr{\nfv,\niv;\ti,\tf}}{\niv\partial\nfv}},
\end{equation}
where the prime at the determinant indicates that the determinant is taken on the $(L-1)\times(L-1)$ submatrix obtained by removing the first line and first row of the full matrix.

The integration over $\thetakj{i}{1}$ is therefore trivial and the final semiclassical expression for the propagator in Fock basis is given by
\begin{equation}
\begin{split}
&K\rbr{\nfv,\niv;\tf,\ti}= \\
&\quad\sul{\gamma:\niv\to\nfv}{}\sqrt{{\det}^\prime\frac{1}{\rbr{-2\pi\rmi\hbar}}\pdiff{^2R_{\gamma}\rbr{\nfv,\niv;\ti,\tf}}{\niv\partial\nfv}}\exp\cbr{\frac{\rmi}{\hbar}R_{\gamma}\rbr{\nfv,\niv;\ti,\tf}},
\end{split}
\label{eq:semiclassical_propagator_Fock}
\end{equation}
where the sum runs now over the \emph{continuous families} of trajectories joining the initial and final occupations.
\section{Coherent backscattering}
\label{sec:cbs}
A natural application of the semiclassical propagator is the transition probability from an initial Fock state $\niv$ to a final one $\nfv$,
\begin{equation}
P\rbr{\nfv,\tf;\niv,\ti}=\abs{\bra{\nfv}\hat{K}\rbr{\tf,\ti}\ket{\niv}}^2=\abs{K\rbr{\nfv,\tf;\niv,\ti}}^2,
\end{equation}
\ie the probability to measure the Fock state $\nfv$ at time $\tf$, if the system was prepared in the Fock state $\niv$ at the initial time $\ti$.

Inserting the semiclassical propagator \eref{eq:semiclassical_propagator_Fock}, yields for the transition probability a double sum over trajectories,
\begin{equation}
P\rbr{\nfv,\tf;\niv,\ti}\approx\sul{\gamma^{},\gamma^\prime:\niv\to\nfv}{}\sqrt{{\det}^\prime\pdiff{^2R_{\gamma^{}}}{\nfv\partial\niv}}\conjg{\sqrt{{\det}^\prime\pdiff{^2R_{\gamma^\prime}}{\nfv\partial\niv}}}\frac{\exp\cbr{\frac{\rmi}{\hbar}\rbr{R_{\gamma^{}}-R_{\gamma^\prime}}}}{\rbr{2\pi\hbar}^{\Sites-1}}.
\label{eq:transition_probability_bosons_double_sum}
\end{equation}
Under disorder average, \ie an average over the on-site energies $\epsilon_l=h_{ll}$, due to the scaling of the action with the total number of particles, the action difference in the exponential gives rise to strong oscillations. Most contributions to the double sum cancel out on disorder average, unless the two actions for trajectories $\gamma^{}$ and $\gamma^\prime$ are correlated.
\subsection{Diagonal approximation}
\begin{figure}
\centering
\includegraphics[width=0.8\textwidth]{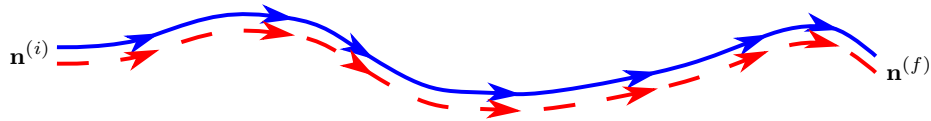}
\put(-325,21){$\niv$}
\put(3,15){$\nfv$}
\caption{\label{fig:da_cbs}Within the diagonal approximation, only identical pairs of trajectories are taken into account}
\end{figure}
The most obvious pairs of trajectories, which are correlated, are the identical ones, \ie $\gamma^{}=\gamma^\prime$ (see \Fref{fig:da_cbs}). For those pairs, the action difference is obviously zero and the transition probability in diagonal approximation reduces to
\begin{equation}
P^{(\mathrm{da})}\rbr{\nfv,\tf;\niv,\ti}=\frac{1}{\rbr{2\pi}^{L-1}}\sul{\gamma}{}\abs{{\det}^\prime\pdiff{\gV{\theta}\rbr{\ti}}{\nfv}},
\end{equation}
where $\gV{\theta}\rbr{\ti}$ is the vector containing the initial phases for the trajectory $\gamma$.

By inserting an integration over the initial phases together with a Dirac delta, ensuring that only those phases contribute which match the ones given by the trajectories and using the properties of the delta function, one can prove the sum rule
\begin{equation}
\begin{split}
\sul{\gamma}{}\abs{{\det}^\prime\pdiff{\gV{\theta}\rbr{\ti}}{\nfv}}&=\intg{0}{2\pi}{^{\Sites-1}\theta^{(i)}}\prodl{\sites=2}{\Sites}\delta\cbr{\abs{\psi_\sites\rbr{\niv;\gV{\theta}^{(i)};\tf}}^2-1/2-\nfj{\sites}} \\
&=\rbr{2\pi}^{\Sites-1}P^{(\mathrm{cl})}\rbr{\nfv,\tf;\niv,\ti},
\end{split}
\label{eq:sum_rule_transition_prob}
\end{equation}
with $\gV{\psi}(\niv;\gV{\theta}^{(i)};\tf)$ being the classical time evolution of the initial state $\gV{\psi}^{(i)}$ with components
\begin{equation}
\psi_\sites^{(i)}=\sqrt{\nij{\sites}+1/2}\exp(\rmi\theta_\sites^{(i)}).
\end{equation}
Therefore, the diagonal approximation yields the classical transition probability
\begin{equation}
P^{(\mathrm{da})}\rbr{\nfv,\tf;\niv,\ti}=P^{(\mathrm{cl})}\rbr{\nfv,\tf;\niv,\ti}.
\end{equation}
\subsection{Coherent backscattering contribution}
\label{subsec:cbs_calculation}
\begin{figure}
\centering
\includegraphics[width=0.6\textwidth]{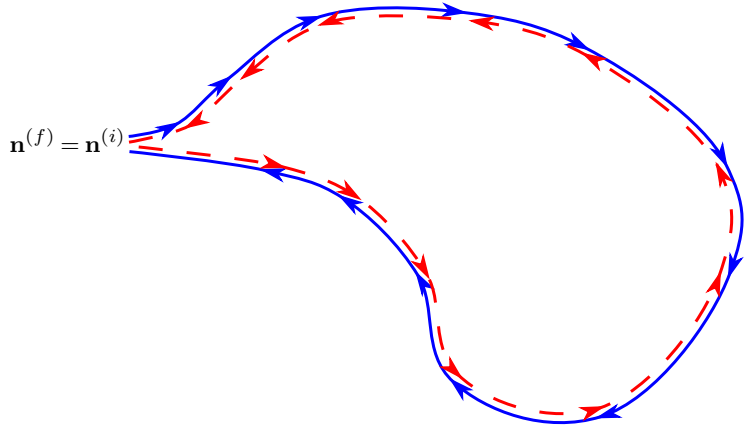}
\put(-275,103){$\nfv=\niv$}
\caption{\label{fig:cbs_cbs}Trajectory pairs responsible for coherent backscattering}
\end{figure}
A second class of correlated pairs of trajectories are those shown in \Fref{fig:cbs_cbs}, namely $\gamma^\prime=\mathcal{T}\gamma^{}$, where $\mathcal{T}$ denotes time reversal. For Bose-Hubbard systems, time reversal is achieved by complex conjugation, such that the time reversal of a solution $\gV{\phi}\rbr{t+\ti}$ of the equations of motion is $\conjg{\gV{\phi}\rbr{\tf-t}}$. Of course, these pairs exist only if the Hamiltonian itself is time reversal symmetric, which for a Bose-Hubbard system corresponds to a real Hamiltonian. It turns out that a Bose-Hubbard chain with nearest neighbor hopping only is always time reversal symmetric, since it can be transformed into a real one by multiplying each component of the wave function with a certain time-independent phase. In the same way, for certain combinations of the phases of the hopping parameter in a Bose-Hubbard ring, the Hamiltonian can be transformed into a real one.

Having identified the time reversed of a trajectory $\gamma$, it is straightforward to check that the actions of both trajectories are the same,
\begin{equation}
R_{\mathcal{T}\gamma}=R_{\gamma}.
\end{equation}
However, in \eref{eq:transition_probability_bosons_double_sum} pairing $\gamma$ with its time reversed is possible only, if $\nfv=\niv$. Therefore, one first has to replace $\gamma$ and $\gamma^\prime$ by nearby trajectories $\tilde{\gamma}$ and $\tilde{\gamma}^\prime$, which both start and end at $\V{n}^{(0)}=\rbr{\niv+\nfv}/2$. In the course of this replacement, one also has to expand the actions in $\nfv$ and $\niv$ up to linear order around $\V{n}^{(0)}$,
\begin{equation}
\begin{split}
&R_{\gamma^{}}-R_{\gamma^{\prime}}\approx \\
&\qquad R_{\tilde{\gamma}^{}}-R_{\tilde{\gamma}^\prime}+\hbar\rbr{\gV{\theta}^{\tilde{\gamma}^{}}\rbr{\tf}-\gV{\theta}^{\tilde{\gamma}^{\prime}}\rbr{\tf}}\cdot\rbr{\nfv-\V{n}^{(0)}}-\hbar\rbr{\gV{\theta}^{\tilde{\gamma}^{}}\rbr{\ti}-\gV{\theta}^{\tilde{\gamma}^{\prime}}\rbr{\ti}}\cdot\rbr{\niv-\V{n}^{(0)}}.
\end{split}
\end{equation}
In the prefactor one can simply replace the second derivative of the action of the original trajectory by the corresponding one of the new trajectory.

If $\tilde{\gamma}^\prime$ is the time reversed of $\tilde{\gamma}$, the initial and final phases of $\tilde{\gamma}^\prime$ are related to those of $\tilde{\gamma}$ by
\begin{align}
\gV{\theta}^{(\tilde{\gamma}^\prime)}\rbr{\tf}=&-\gV{\theta}^{(\tilde{\gamma})}\rbr{\ti} \\
\gV{\theta}^{(\tilde{\gamma}^\prime)}\rbr{\ti}=&-\gV{\theta}^{(\tilde{\gamma})}\rbr{\tf},
\end{align}
which, after applying the sum rule \eref{eq:sum_rule_transition_prob}, yields for the coherent backscattering contribution to the transition probability
%
%\begin{figure}
%\centering
%\includegraphics[width=0.8\textwidth]{cbs/figs/fivesites_new.pdf}
%\caption{\label{fig:cbs_numerics_timedep}Emergence of the coherent backscattering for a Bose-Hubbard chain with five sites from numerical calculations. The initial occupation was chosen to be $(3,2,3,4,2)$. The red crosses are the classical transition probability, while the black diamonds are the quantum ones. The evolution times are (a) $t=1.5\hbar/J$, (b) $t=2.5\hbar/J$, (c) $t=5\hbar/J$, (d) $t=10\hbar/J$, (e) $t=20\hbar/J$, (f) $t=50\hbar/J$, where $J$ is the hopping parameter. Figure by courtesy of Julien Dujardin, Arturo Arg\"uelles and Peter Schlagheck.}
%\end{figure}
%
\begin{equation}
\begin{split}
P^{(\mathrm{cbs})}&\rbr{\nfv,\tf;\niv,\tf}= \\
&\frac{\delta_{TRI}}{\rbr{2\pi}^{\Sites-1}}\intg{0}{2\pi}{^{\Sites-1}\theta^{(i)}}\exp\tbr{\rmi\cbr{\gV{\theta}^{(i)}+\gV{\theta}\rbr{\niv,\gV{\theta}^{(i)};\tf}}\cdot\rbr{\nfv-\niv}} \\
&\pushright{\times\prodl{\sites=2}{\Sites}\delta\cbr{\abs{\psi_l\rbr{\niv,\thetakv{i};\tf}}^2-\frac{1}{2}-\niv},}
\end{split}
\end{equation}
where $\delta_{TRI}$ is one, if the system is time reversal invariant and zero otherwise.

For a classically chaotic system, within a disorder average one can replace the final phases $\gV{\theta}(\niv,\gV{\theta}^{(i)};\tf)$ by identically, independent, in $\cbr{0,2\pi}$ uniformly distributed random variables, such that for $P^{(cbs)}$ one can perform an average over the final phases yielding a factor $\delta_{\niv,\nfv}$. Thus, the averaged coherent backscattering contribution to the transition probability is given by
\begin{equation}
P^{(\mathrm{cbs})}\rbr{\nfv,\tf;\niv,\ti}=\delta_{TRI}\delta_{\niv,\nfv}P^{(\mathrm{cl})}\rbr{\nfv,\tf;\niv,\ti}.
\end{equation}

As long as time reversal symmetry is the only discrete symmetry of the system, pairing a trajectory with itself or its time reverse, are the only two generic possibilities to get a contribution with vanishing action difference. All further possible pairs of trajectories give contributions which are suppressed by a factor of $1/\ntot$ compared to the two latter. In the semiclassical limit, the averaged transition probability is then given by
\begin{equation}
P\rbr{\nfv,\tf;\niv,\ti}=\rbr{1+\delta_{TRI}\delta_{\nfv,\niv}}P^{(\mathrm{cl})}\rbr{\nfv,\tf;\niv,\ti},
\label{eq:cbs_bosons}
\end{equation}
\ie the quantum transition probability is enhanced for $\nfv=\niv$ compared to the classical one by a factor of two due to the constructive interference between time-reversed paths.
%
%\begin{figure}
%\centering
%\includegraphics[width=0.9\textwidth]{cbs/figs/newsixring.pdf}
%\caption{\label{fig:cbs_numerics_low_occupations}Dependence of the coherent backscattering peak on the phase $\phi$ of the hopping parameter $J$ for a Bose-Hubbard ring consisting of six sites (see inset of (a)). The initial occupations were chosen to be (a) $(3,3,2,3,4,2)$ and $(1,1,2,2,1,0)$. Figure by courtesy of Julien Dujardin, Arturo Arg\"uelles and Peter Schlagheck.}
%\end{figure}

However, only paths which are not self-retracing may contribute to the coherent backscattering peak, because the time reverse of a self-retracing path is again the path itself and therefore this pairing would already be included in the diagonal approximation. For short times, one can expect that all trajectories coming back to the initial occupations are all self-retracing and therefore a certain minimal time is required to observe coherent backscattering (see supplementary material of \cite{cbs_fock}). This behavior as well as the existence of coherent backscattering itself has been shown numerically in \cite{cbs_fock}.

We remark that, by its very definition, coherent backscattering in Fock space requires a dephasing mechanism (in our case the combined effect of dynamical instability and average over disorder realizations) and it is therefore a robust effect against external perturbations as long as they do not produce decoherence. The existence of coherent but robust interference effects in many-body systems is the more remarkable considering the extreme delicacy of the associated quantum coherences. Also, the prediction and numerical confirmation of this dynamical echo effect has deep implications in the very active field of thermalisation of closed, interacting quantum systems (see \cite{therm} for a recent review). There, the main question is whether local observables of many-body systems reach an equilibrium value in agreement with the expectations from quantum statistical physics. Our findings show that, even under strong ensemble average (which is expected to speed up thermalisation) if the system preserves time reversal symmetry, the memory of the many-body initial state remains for arbitrarily long times. Although this observation is not in contrast with thermalisation as understood in \cite{therm}, as the effect of coherent backscattering on the expectation values of {\it local} observables is a $1/N$ effect, it indicates that time-reversal may play an unexpected role in the thermal equilibration of many-body systems.
%is shown in \Fref{fig:cbs_numerics_timedep}.
%
%\Fref{fig:cbs_numerics_low_occupations} shows that the peak indeed vanishes, when time reversal invariance is broken, which clearly identifies the source of the peak to be coherent backscattering. The fact that for certain phases $\phi$ (apart from $2\pi$) of the hopping parameter the coherent backscattering peak reappears is due to the effective reestablishment of time reversal invariance (see \cite{cbs_fock}). Moreover, \Fref{fig:cbs_numerics_low_occupations} shows that -- although the semiclassical theory is strictly speaking valid for large occupations only, coherent backscattering can be observed already for very small particle number.
%
%
%
%
\section{Conclusion}
\label{sec:conclusion}
In this paper, we have rigorously constructed an approximation for the quantum mechanical transition amplitude connecting two many-body states of a system of interacting bosons. We applied this method to predict an echo type backscattering effect due to quantum interference in Fock space. Our approximation is semiclassical in spirit, in the sense that it relies on a stationary phase analysis of the exact path integral representation of the quantum
mechanical propagator, but corresponds to an asymptotic, classical limit that is different from the usual $\hbar \to 0$ limit . Rather, we show that when written in the language of fields, an interacting bosonic system has as classical limit a non-linear classical field theory, namely the mean-field Gross-Pitaevskii equation. The semiclassical method then unambiguously prescribes how to use the solutions of this equation to describe quantum processes in terms of the actions and stabilities of the classical solutions, and in this way we find that the mean-field limit, commonly used only in the context of ground-state properties due to its usual derivation from a variational approach, actually contains all the information one needs to understand quantum mechanical effects for excited states, including quantum fluctuations and interference. This conceptual step is one of the main results of this paper.

Once the semiclassical propagator is derived and its domain of validity is clarified, we turned to its application to give a semiclassical picture of many-body interference in Fock space. Coherent sums over semiclasssical amplitudes then lead to constructive interference of the transition probability that is not captured in any mean-field level of approximation. Following the steps learned from the study of quantum interference in mesoscopic systems, we then calculate the Fock-space version of the mesoscopic dynamical echo, and found a universal enhancement of the return probability due to the interplay between interference and interactions. This prediction,
successfully confirmed in numerical calculations presented in \cite{cbs_fock} shows the power of the semiclassical thinking in many-body system, a road that has shown to be useful for other setups like bosonic transport in optical lattices \cite{TomPet}, connecting soliton-like solutions of discrete non-linear equations with properties of the quantum spectra \cite{Trace}, and going beyond the classical Truncated Wigner method to describe dynamical processes in many-body systems \cite{TomPet2}. Further related questions such as the extension to the continuum limit necessary to deal with scattering problems and the extension to Fermionic fields (partially initiated in \cite{fermionic_propagator}) are subject of work in progress.
\enlargethispage{20pt}
%
%
%\ethics{Insert ethics statement here if applicable.}
%
%\dataccess{Insert details of how to access any supporting data here.}
%
%\aucontribute{For manuscripts with two or more authors, insert details of the authors’ contributions here. This should take the form: 'AB caried out the experiments. CD performed the data analysis. EF conceived of and designed the study, and drafted the manuscript All auhtors read and approved the manuscript'.}
%
\competing{The authors declare that they have no competing interests.}
\funding{This work was financially supported by the Deutsche Forschungsgemeinschaft within the research unit FOR 760.}
\ack{The authors thank Peter Schlagheck for attendance, cooperation and useful conversation and discussion during the whole course of this project.}

%\disclaimer{Insert disclaimer text here if applicable.}
%
%
%
%
%
\begin{appendix}
\section{The path integral in quadrature representation}
\label{app:path-integral}
\begin{equation}
\begin{split}
\Braket{\qkv{f}|\hat{K}\rbr{\tf,\ti}|\qkv{i}}=&\pilimit\Braket{\qkv{f}|\prodl{\picount=1}{\pisteps}\exp\cbr{-\frac{\rmi\timestep}{\hbar}\hat{H}\rbr{\tf-\picount\timestep}}|\qkv{i}} \\
=&\pilimit\intg{}{}{^\Sites\qkv{1}}\cdots\intg{}{}{^\Sites\qkv{\pisteps-1}}\prodl{\picount=1}{\pisteps}\Braket{\qkv{\picount}|\exp\cbr{-\frac{\rmi\timestep}{\hbar}\hat{H}\rbr{\ti+\picount\timestep}}|\qkv{\picount-1}}.
\end{split}
\label{eq:path-integral_q-only}
\end{equation}
In order to evaluate the matrix elements of the right hand side of \Eref{eq:path-integral_q-only}, a Hubbard-Stratonovich transformation has to be performed,
\begin{equation}
\begin{split}
&\exp\rbr{-\frac{\rmi\timestep}{2\hbar}\sul{\sites^{},\sites^{\prime},\sites^{\prime\prime},\sites^{\prime\prime\prime}}{}\hat{a}^{\dagger}_{\sites^{}}\hat{a}^{}_{\sites^{\prime}}U_{\sites^{}\sites^{\prime}\sites^{\prime\prime}\sites^{\prime\prime\prime}}\hat{a}^{\dagger}_{\sites^{\prime\prime}}\hat{a}^{}_{\sites^{\prime\prime\prime}}} \\
&\qquad=\frac{\displaystyle\intg{}{}{\sigma}\exp\cbr{\frac{\rmi\timestep}{2\hbar}\sul{\sites^{},\sites^{\prime},\sites^{\prime\prime},\sites^{\prime\prime\prime}}{}\sigma_{\sites^{}\sites^{\prime}}\rbr{U^{-1}}_{\sites^{}\sites^{\prime}\sites^{\prime\prime}\sites^{\prime\prime\prime}}\sigma_{\sites^{\prime\prime}\sites^{\prime\prime\prime}}-\frac{\rmi\timestep}{\hbar}\sul{\sites^{}\sites^{\prime}}{}\sigma_{\sites^{}\sites^{\prime}}\hat{a}^{\dagger}_{\sites^{}}\hat{a}^{}_{\sites^{\prime}}}}{\displaystyle\intg{}{}{\sigma}\exp\cbr{\frac{\rmi\timestep}{2\hbar}\sul{\sites^{},\sites^{\prime},\sites^{\prime\prime},\sites^{\prime\prime\prime}}{}\sigma_{\sites^{}\sites^{\prime}}\rbr{U^{-1}}_{\sites^{}\sites^{\prime}\sites^{\prime\prime}\sites^{\prime\prime\prime}}\sigma_{\sites^{\prime\prime}\sites^{\prime\prime\prime}}}},
\end{split}
\label{eq:hubbard-stratonovich}
\end{equation}
where the integration runs over hermitian matrices $\sigma$ and $\rbr{U^{-1}}_{klmn}$ is chosen such that $\sul{a,b}{}\rbr{U^{-1}}_{kalb}U_{ambn}=U_{akbl}\rbr{U^{-1}}_{ambn}=\delta_{km}\delta_{ln}$. This transformation can be proven by shifting $\sul{a,b}{}\sigma_{kl}\to\sigma_{kl}+U_{kalb}\hat{a}^{\dagger}_{a}\hat{a}^{}_{b}$ and using the relations $U_{klmn}^\ast=U_{mnkl}$ and $U_{klmn}=U_{lkmn}=U_{klnm}$ induced by the hermiticity of the Hamiltonian \eref{eq:general_BH}.

Applying \Eref{eq:hubbard-stratonovich} to the matrix elements in \Eref{eq:path-integral_q-only} for a Hamiltonian of the form \Eref{eq:general_BH} yields
\begin{equation}
\begin{split}
&\Braket{\qkv{\picount}|\exp\cbr{-\frac{\rmi\timestep}{\hbar}\hat{H}\rbr{\ti+\picount\timestep}}|\qkv{\picount-1}}= \\
&\frac{\displaystyle\intg{}{}{\sigma}\exp\cbr{\frac{\rmi\timestep}{2\hbar}\sul{\sites^{},\sites^{\prime},\sites^{\prime\prime},\sites^{\prime\prime\prime}}{}\sigma_{\sites^{}\sites^{\prime}}\rbr{U^{-1}}_{\sites^{}\sites^{\prime}\sites^{\prime\prime}\sites^{\prime\prime\prime}}\sigma_{\sites^{\prime\prime}\sites^{\prime\prime\prime}}}\Braket{\qkv{\picount}|\exp\cbr{-\frac{\rmi\timestep}{\hbar}\sul{\sites^{}\sites^{\prime}}{}\rbr{h^{(\picount)}_{\sigma}}_{\sites^{}\sites^{\prime}}\hat{a}^{\dagger}_{\sites^{}}\hat{a}^{}_{\sites^{\prime}}}|\qkv{\picount-1}}}{\displaystyle\intg{}{}{\sigma}\exp\cbr{\frac{\rmi\timestep}{2\hbar}\sul{\sites^{},\sites^{\prime},\sites^{\prime\prime},\sites^{\prime\prime\prime}}{}\sigma_{\sites^{}\sites^{\prime}}\rbr{U^{-1}}_{\sites^{}\sites^{\prime}\sites^{\prime\prime}\sites^{\prime\prime\prime}}\sigma_{\sites^{\prime\prime}\sites^{\prime\prime\prime}}}},
\end{split}
\label{eq:path-integral_operators_quadratic}
\end{equation}
with $h_{\sigma}$ being the effective single-particle hamiltonian matrix with entries given by
\begin{equation}
\rbr{h_{\sigma}^{(\picount)}}_{\sites^{}\sites^\prime}=h_{\sites^{}\sites^{\prime}}\rbr{\ti+\picount\timestep}-\frac{1}{2}\sul{\sites^{\prime\prime}}{}U_{\sites^{}\sites^{\prime\prime}\sites^{\prime\prime}\sites^{\prime}}\rbr{\ti+\picount\timestep}+\sigma_{\sites^{}\sites^{\prime}}.
\end{equation}
Next, after replacing $\hat{a}_{\sites}^{\dagger}$ and $\hat{a}_{\sites}^{}$ by $\hat{q}_{\sites}$ and $\hat{p}_{\sites}$, one has to replace the exponential containing $\hat{q}$ \emph{and} $\hat{p}$ operators by a product of exponentials, which contain only $\hat{q}$ or only $\hat{p}$ operators. The product has to be arranged such that the $\hat{q}$ operators can be applied directly to the quadrature states at the right or at the left. To this end recognize that
\begin{equation}
\sul{\sites^{}\sites^{\prime}}{}\rbr{h^{(\picount)}_{\sigma}}_{\sites^{}\sites^{\prime}}\rbr{\hat{q}_{\sites^{}}-\rmi\hat{p}_{\sites^{}}}\rbr{\hat{q}_{\sites^{\prime}}+\rmi\hat{p}_{\sites^{\prime}}}=\sul{\sites^{}\sites^{\prime}}{}\cbr{\rbr{h^{(\picount)}_{\sigma}}_{\sites^{}\sites^{\prime}}\rbr{\hat{q}_{\sites^{}}\hat{q}_{\sites^{\prime}}+\hat{p}_{\sites^{}}\hat{p}_{\sites^{\prime}}-2\quadfac^2\delta_{\sites^{}\sites^{\prime}}}-\Im\rbr{h_{\sigma}^{(\picount)}}_{\sites^{}\sites^{\prime}}\rbr{\hat{q}_{\sites^{}}+\hat{p}_{\sites^{}}}\rbr{\hat{q}_{\sites^{\prime}}+\hat{p}_{\sites^{\prime}}}}.
\label{eq:prepare_for_splitting}
\end{equation}
Note here that $\Im\rbr{h_{\sigma}^{(\picount)}}$ is antisymmetric and thus its diagonal is zero.

The last term in \Eref{eq:prepare_for_splitting} still terms $\hat{q}_{\sites^{}}\hat{p}_{\sites^{\prime}}$. Thus before splitting up the exponentials, we need to get rid of these terms. This is possible by means of a method known as ``uncompleting the square'' or ``Gaussian trick'',

\begin{equation}
\exp\cbr{\frac{\rmi\timestep}{4\quadfac^2\hbar}\sul{\sites^{},\sites^{\prime}}{}\Im\rbr{h_{\sigma}^{(\picount)}}_{\sites^{}\sites^{\prime}}\rbr{\hat{q}_{\sites^{}}+\hat{p}_{\sites^{}}}\rbr{\hat{q}_{\sites^{\prime}}+\hat{p}_{\sites^{\prime}}}}=\int\frac{\rmd^{\Sites}u}{\sqrt{2\pi}^{\Sites}}\exp\cbr{-\frac{u^2}{2}-\sqrt{\frac{\tau}{2\quadfac^2\hbar}}\sul{\sites^{},\sites^{\prime}}{}\cbr{\sqrt{\Im\rbr{h_{\sigma}^{(m)}}}}_{\sites^{}\sites^{\prime}}u_{\sites^{}}\rbr{\hat{q}_{\sites^{\prime}}+\hat{p}_{\sites^{\prime}}}}.
\label{eq:splitting_mixed}
\end{equation}

When splitting up the exponential, one should recognize that not only the last term in \Eref{eq:splitting_mixed} is proportional to $\sqrt{\tau}$, but due to the Gaussian weight of the integration $\sigma\sim1/\sqrt{\tau}$. Thus by applying
\begin{equation}
\exp\cbr{\lambda\rbr{\hat{A}+\hat{B}}}=\exp\rbr{\frac{\lambda}{2}A}\exp\rbr{\lambda B}\exp\rbr{\frac{\lambda}{2}A}+\mathcal{O}\rbr{\lambda^3}
\end{equation}
terms proportional to $\tau$ are correctly taken into account and yields for the matrix elements in \Eref{eq:path-integral_operators_quadratic} after also introducing a unit operator in terms of $\Ket{\V{p}}\Bra{\V{p}}$
\begin{equation}
\begin{split}
&\Braket{\qkv{\picount}|\exp\cbr{-\frac{\rmi\timestep}{\hbar}\sul{\sites^{}\sites^{\prime}}{}\rbr{h^{(\picount)}_{\sigma}}_{\sites^{}\sites^{\prime}}\hat{a}^{\dagger}_{\sites^{}}\hat{a}^{}_{\sites^{\prime}}}|\qkv{\picount-1}} \\
&=\intg{}{}{^\Sites\pkj{\picount}{}}\int\frac{\rmd^\Sites u}{\sqrt{2\pi}^\Sites}\exp\rbr{-\frac{\V{u}^2}{2}+\frac{\rmi\timestep}{2\hbar}\Tr h_{\sigma}^{(\picount)}} \\
&\qquad\times\left\langle\qkv{\picount}\left|\exp\tbr{-\sul{\sites^{},\sites^{\prime}}{}\cbr{\frac{\rmi\timestep}{8\quadfac^2\hbar}\rbr{h_{\sigma}^{(\picount)}}_{\sites^{}\sites^{\prime}}\hat{q}_{\sites^{}}\hat{q}_{\sites^{\prime}}-\sqrt{\frac{\timestep}{8\quadfac^2\hbar}}\Im\rbr{h_{\sigma}^{(\picount)}}_{\sites^{}\sites^{\prime}}u_{\sites^{}}\hat{q}_{\sites^{\prime}}}}\right.\right. \\
&\qquad\qquad\qquad\,\left.\left.\times\exp\tbr{-\sul{\sites^{},\sites^{\prime}}{}\cbr{\frac{\rmi\timestep}{4\quadfac^2\hbar}\rbr{h_{\sigma}^{(\picount)}}_{\sites^{}\sites^{\prime}}\hat{p}_{\sites^{}}\hat{p}_{\sites^{\prime}}-\sqrt{\frac{\timestep}{2\quadfac^2\hbar}}\Im\rbr{h_{\sigma}^{(\picount)}}_{\sites^{}\sites^{\prime}}u_{\sites^{}}\hat{p}_{\sites^{\prime}}}}\right|\pkv{\picount}\right\rangle \\
&\qquad\times\Braket{\pkv{\picount}|\exp\tbr{-\sul{\sites^{},\sites^{\prime}}{}\cbr{\frac{\rmi\timestep}{8\quadfac^2\hbar}\rbr{h_{\sigma}^{(\picount)}}_{\sites^{}\sites^{\prime}}\hat{q}_{\sites^{}}\hat{q}_{\sites^{\prime}}-\sqrt{\frac{\timestep}{8\quadfac^2\hbar}}\Im\rbr{h_{\sigma}^{(\picount)}}_{\sites^{}\sites^{\prime}}u_{\sites^{}}\hat{q}_{\sites^{\prime}}}}|\qkv{\picount-1}}
\end{split}
\end{equation}
Now, the operators $\hat{p}_{\sites}$ can simply be replaced by $\qkv{\picount}{\sites}$ and the operators $\hat{q}_{\sites}$ in the most left exponential by $\qkv{\picount}{\sites}$, while those in the right exponential can be replaced by $\qkj{\picount-1}{\sites}$. After that, first the integration over $\V{u}$ and finally the one over $\sigma$ can be carried out, thus undoing the beforehand performed transformations, yielding
\begin{align}
&\Braket{\qkv{\picount}|\exp\cbr{-\frac{\rmi\timestep}{\hbar}\hat{H}\rbr{\ti+\picount\timestep}}|\qkv{\picount-1}}= \\
&\quad\int\frac{\rmd^{\Sites}\pkj{\picount}{}}{\rbr{4\pi\quadfac^2}^{\Sites}}\exp\tbr{\frac{\rmi\timestep}{2\quadfac^2}\pkv{\picount}\cdot\rbr{\qkv{\picount}-\qkv{\picount-1}}-\tilde{H}^{(cl)}\rbr{\psikvd{\picount},\psikv{\picount};\left.{\tilde{\gV{\psi}}^{(\picount)}}\right.^\ast,\tilde{\gV{\psi}}^{(\picount};\ti+\picount\timestep}}, \nonumber
\end{align}
where $\psikv{\picount}=\rbr{\qkv{\picount}+\rmi\pkv{\picount}}/\rbr{2\quadfac}$, $\tilde{\gV{\psi}}^{(\picount)}=\rbr{\qkv{\picount-1}+\rmi\pkv{\picount}}/\rbr{2\quadfac}$ and
\begin{align}
&\tilde{H}^{(cl)}\rbr{\psikvd{\picount},\psikv{\picount};\left.{\tilde{\gV{\psi}}^{(\picount)}}\right.^\ast,\tilde{\gV{\psi}}^{(\picount};\ti+\picount\timestep}= \\
&\qquad\sul{\sites^{}\sites^{\prime}}{}\rbr{h_{\sites^{}\sites^{\prime}}\rbr{\ti+\picount\timestep}-\frac{1}{2}\sul{\sites^{\prime\prime}}{}U_{\sites^{}\sites^{\prime\prime}\sites^{\prime\prime}\sites^{\prime}}\rbr{\ti+\picount\timestep}}\frac{\left.\psi_{\sites^{}}^{(\picount)}\right.^\ast\psi_{\sites^{\prime}}^{(\picount)}+\left.\tilde{\psi}_{\sites^{}}^{(\picount)}\right.^\ast\tilde{\psi}_{\sites^{\prime}}^{(\picount)}-\delta_{\sites^{}\sites^{\prime}}}{2} \nonumber \\
&\qquad+\frac{1}{2}\sul{\sites^{}\sites^{\prime}\sites^{\prime\prime}\sites^{\prime\prime\prime}}{}U_{\sites^{}\sites^{\prime}\sites^{\prime\prime}\sites{\prime\prime\prime}}\frac{\left.\psi_{\sites^{}}^{(\picount)}\right.^\ast\psi_{\sites^{\prime\prime}}^{(\picount)}+\left.\tilde{\psi}_{\sites^{}}^{(\picount)}\right.^\ast\tilde{\psi}_{\sites^{\prime\prime}}^{(\picount)}-\delta_{\sites^{}\sites^{\prime\prime}}}{2}\frac{\left.\psi_{\sites^{\prime}}^{(\picount)}\right.^\ast\psi_{\sites^{\prime\prime\prime}}^{(\picount)}+\left.\tilde{\psi}_{\sites^{\prime}}^{(\picount)}\right.^\ast\tilde{\psi}_{\sites^{\prime\prime\prime}}^{(\picount)}-\delta_{\sites^{\prime}\sites^{\prime\prime\prime}}}{2} \nonumber
\end{align}
Note that in the continuous limit, $\tau\to0$, $\tilde{\gV{\psi}}^{(\picount)},\psikv{\picount}\to\gV{\psi}\rbr{t}$ and therefore
\begin{equation}
\tilde{H}^{(cl)}\rbr{\psikvd{\picount},\psikv{\picount};\left.{\tilde{\gV{\psi}}^{(\picount)}}\right.^\ast,\tilde{\gV{\psi}}^{(\picount};\ti+\picount\timestep}\to\Hcl\rbr{\gVd{\psi}\rbr{t},\gV{\psi}\rbr{t};t},
\end{equation}
where $\Hcl$ is given by \Eref{eq:classical_limit_BH}.
\section{The semiclassical prefactor in quadrature representation}
\label{app:scl_prefactor_quad}
In order to derive the semiclassical prefactor, \Eref{eq:semiclassical_prefactor_quad}, according to the stationary phase approximation, the second variation of the exponent around the classical trajectory is kept for the integration yielding gaussian integrals,
\begin{align}
\mathcal{A}_{\gamma}=&\pilimit\frac{1}{\rbr{4\pi\quadfac^2}^{\pisteps L}}\intg{}{}{^L\delta\qkj{1}{}}\cdots\intg{}{}{^L\delta\qkj{\pisteps-1}{}}\intg{}{}{^L\delta\pkj{1}{}}\cdots\intg{}{}{^L\delta\pkj{\pisteps}{}}
\label{eq:semiclassical_prefactor_quad_start} \\
&\exp\Bigg\{\frac{\rmi}{2\quadfac^2}\delta\pkv{1}\cdot\delta\qkv{1}+\frac{\rmi}{2\quadfac^2}\sul{\picount=2}{\pisteps-1}\delta\pkv{\picount}\cdot\rbr{\delta\qkv{\picount}-\delta\qkv{\picount-1}}-\frac{\rmi}{2\quadfac^2}\delta\pkv{\pisteps}\cdot\delta\qkv{\pisteps-1} \nonumber \\
&-\frac{\rmi\timestep}{\hbar}\sul{\picount=1}{\pisteps-1}\left(\begin{array}{c} \delta\qkv{\picount} \\ \delta\pkv{\picount} \end{array}\right)\pdiff{^2\tilde{H}^{(cl)}\rbr{\psikvd{\picount},\psikv{\picount};\left.{\tilde{\gV{\psi}}^{(\picount)}}\right.^\ast,\tilde{\gV{\psi}}^{(\picount};\ti+\picount\timestep}}{\left(\delta\qkv{\picount},\delta\pkv{\picount}\right)^2}\left(\begin{array}{c} \delta\qkv{\picount} \\ \delta\pkv{\picount} \end{array}\right) \nonumber \\
&-\frac{\rmi\timestep}{\hbar}\sul{\picount=2}{\pisteps}\left(\begin{array}{c} \delta\qkv{\picount-1} \\ \delta\pkv{\picount} \end{array}\right)\pdiff{^2\tilde{H}^{(cl)}\rbr{\psikvd{\picount},\psikv{\picount};\left.{\tilde{\gV{\psi}}^{(\picount)}}\right.^\ast,\tilde{\gV{\psi}}^{(\picount};\ti+\picount\timestep}}{\left(\delta\qkv{\picount-1},\delta\pkv{\picount}\right)^2}\left(\begin{array}{c} \delta\qkv{\picount-1} \\ \delta\pkv{\picount} \end{array}\right)\Bigg\}, \nonumber % \\
%&\qquad-\frac{\rmi\timestep}{\hbar}\delta\pkv{1}\pdiff{^2\Hcl\rbr{\psikvd{1},\psikv{1};\ti+\timestep}}{{\delta\pkv{1}}^2}\delta\pkv{1}\Bigg\}, \nonumber
\end{align}
which after substituting $\xkv{\picount}=2\quadfac\V{x}^{(\picount)}$, $\ykv{\picount}=2\quadfac\V{y}^{(\picount)}$ and introducing the short hand notation $H^{(\picount)}=\tilde{H}^{(cl)}\rbr{\psikvd{\picount},\psikv{\picount};\left.{\tilde{\gV{\psi}}^{(\picount)}}\right.^\ast,\tilde{\gV{\psi}}^{(\picount};\ti+\picount\timestep}$ reads
\begin{align}
&\mathcal{A}_\gamma= \\
&\pilimit\intg{}{}{^\Sites x^{(1)}{}}\cdots\intg{}{}{^\Sites x^{(\pisteps-1)}{}}\intg{}{}{^\Sites y^{(1)}{}}\cdots\intg{}{}{^\Sites y^{(\pisteps)}{}}\frac{1}{\pi^{\pisteps \Sites}\rbr{2\quadfac}^\Sites} \nonumber \\
&\times\exp\Bigg\{\rmi\sul{\picount=1}{\pisteps-1}\Bigg[\rbr{\begin{array}{c}\V{x}^{(\picount)}\\\V{y}^{(\picount)}\end{array}}\rbr{\begin{array}{cc}-\frac{2\quadfac^2\timestep}{\hbar}\pdiff{^2\rbr{H^{(\picount+1)}+H^{(m)}}}{{\qkv{\picount}}^2}&\unity{\Sites}-\frac{2\quadfac^2\timestep}{\hbar}\pdiff{^2H^{(m)}}{\qkv{\picount}\partial\pkv{\picount}}\\\unity{\Sites}-\frac{2\quadfac^2\timestep}{\hbar}\pdiff{^2H^{(m)}}{\pkv{\picount}\partial\qkv{\picount}}&-\frac{2\quadfac^2\timestep}{\hbar}\pdiff{^2H^{(\picount)}}{{\pkv{\picount}}^2}\end{array}}\rbr{\begin{array}{c}\V{x}^{(\picount)}\\\V{y}^{(\picount)}\end{array}} \nonumber \\
&\qquad\qquad-2\rmi\rbr{\begin{array}{c}\V{x}^{(\picount)}\\\V{y}^{(\picount)}\end{array}}\rbr{\begin{array}{cc}\frac{2\quadfac^2\timestep}{\hbar}\pdiff{^2H^{(\picount+1)}}{\qkv{\picount}\partial\qkv{\picount+1}}&\unity{\Sites}+\frac{2\quadfac^2\timestep}{\hbar}\pdiff{^2H^{(\picount+1)}}{\qkv{\picount}\partial\pkv{\picount+1}}\\0&0\end{array}}\rbr{\begin{array}{c}\V{x}^{(\picount+1)}\\\V{y}^{(\picount+1)}\end{array}}\Bigg] \nonumber \\
&\qquad\qquad-\frac{2\rmi\quadfac^2\timestep}{\hbar}\V{y}^{(\pisteps)}\pdiff{^2H^{(\pisteps)}}{{\pkv{\pisteps}}^2}\V{y}^{(\pisteps)}\Bigg\}, \nonumber
\end{align}
where $\unity{\Sites}$ is the $\Sites\times\Sites$ unit matrix and the result has to be evaluated at $\xkv{\pisteps}=0$.

For the evaluation of these integrals we define the matrices $V^{(\picount)},W^{(\picount)}$ and $Y^{(\picount)}$, such that after integrating $\xkv{1},\ykv{1},\ldots\xkv{\picount^{\prime}-1},\ykv{\picount^{\prime}-1}$, the semiclassical amplitude is given by
\begin{align}
&\mathcal{A}_\gamma=
\label{eq:semiclassical_amplitude_mprime-integrals} \\
&\pilimit\intg{}{}{^\Sites x^{(\picount^{\prime})}{}}\cdots\intg{}{}{^\Sites x^{(\pisteps-1)}{}}\intg{}{}{^\Sites y^{(\picount^{\prime})}{}}\cdots\intg{}{}{^\Sites y^{(\pisteps)}{}}\frac{\sqrt{\prodl{\picount=1}{\picount^{\prime}-1}\det\rbr{\begin{array}{cc} V^{(\picount)} & W^{(\picount)} \\ \left.W^{(\picount)}\right.^{\rm T} & Y^{(\picount)} \end{array}}}}{\pi^{\rbr{\pisteps-\picount^{\prime}+1} \Sites}\rbr{2\quadfac}^\Sites} \nonumber \\
&\times\exp\Bigg\{\rmi\rbr{\begin{array}{c}\V{x}^{(\picount^{\prime})}\\\V{y}^{(\picount^{\prime})}\end{array}}\rbr{\begin{array}{cc} V^{(\picount^{\prime})} & W^{(\picount^{\prime})} \\ \left.W^{(\picount^{\prime})}\right.^{\rm T} & Y^{(\picount^{\prime})} \end{array}}\rbr{\begin{array}{c}\V{x}^{(\picount^{\prime})}\\\V{y}^{(\picount^{\prime})}\end{array}} \nonumber \\
&+\rmi\sul{\picount=\picount^{\prime}+1}{\pisteps-1}\rbr{\begin{array}{c}\V{x}^{(\picount)}\\\V{y}^{(\picount)}\end{array}}\rbr{\begin{array}{cc}-\frac{2\quadfac^2\timestep}{\hbar}\pdiff{^2\rbr{H^{(\picount+1)}+H^{(m)}}}{{\qkv{\picount}}^2}&\unity{\Sites}-\frac{2\quadfac^2\timestep}{\hbar}\pdiff{^2H^{(m)}}{\qkv{\picount}\partial\pkv{\picount}}\\\unity{\Sites}-\frac{2\quadfac^2\timestep}{\hbar}\pdiff{^2H^{(m)}}{\pkv{\picount}\partial\qkv{\picount}}&-\frac{2\quadfac^2\timestep}{\hbar}\pdiff{^2H^{(\picount)}}{{\pkv{\picount}}^2}\end{array}}\rbr{\begin{array}{c}\V{x}^{(\picount)}\\\V{y}^{(\picount)}\end{array}} \nonumber \\
&-2\rmi\sul{\picount=\picount^{\prime}}{\pisteps-1}\rbr{\begin{array}{c}\V{x}^{(\picount)}\\\V{y}^{(\picount)}\end{array}}\rbr{\begin{array}{cc}\frac{2\quadfac^2\timestep}{\hbar}\pdiff{^2H^{(\picount+1)}}{\qkv{\picount}\partial\qkv{\picount+1}}&\unity{\Sites}+\frac{2\quadfac^2\timestep}{\hbar}\pdiff{^2H^{(\picount+1)}}{\qkv{\picount}\partial\pkv{\picount+1}}\\0&0\end{array}}\rbr{\begin{array}{c}\V{x}^{(\picount+1)}\\\V{y}^{(\picount+1)}\end{array}} \nonumber \\
&\qquad\qquad-\frac{2\rmi\quadfac^2\timestep}{\hbar}\V{y}^{(\pisteps)}\pdiff{^2H^{(\pisteps)}}{{\pkv{\pisteps}}^2}\V{y}^{(\pisteps)}\Bigg\}, \nonumber
\end{align}
Obviously, $V^{(\picount)}$, $W^{(\picount)}$, $Y^{(\picount)}$ fulfill the initial conditions
\begin{equation}
\begin{split}
V^{(1)}&=-\frac{2\quadfac^2\timestep}{\hbar}\pdiff{^2\rbr{H^{(1)}+H^{(2)}}}{\left.\qkv{1}\right.^2}, \\
W^{(1)}&=\unity{\Sites}-\frac{2\quadfac^2\timestep}{\hbar}\pdiff{^2H^{(1)}}{\qkv{1}\partial\pkv{1}}, \\
Y^{(1)}&=-\frac{2\quadfac^2\timestep}{\hbar}\pdiff{^2H^{(1)}}{\left.\pkv{1}\right.^2}.
\end{split}
\end{equation}
Performing the integrals over $\xkv{\picount^{\prime}}$ and $\ykv{\picount^{\prime}}$ in \Eref{eq:semiclassical_amplitude_mprime-integrals} and using inverse of a matrix in block form,
\begin{equation}
\begin{split}
\left(\begin{array}{cc} A & B \\ C & D \end{array}\right)^{-1}&=\left(\begin{array}{cc} \rbr{A-BD^{-1}C}^{-1} & -A^{-1}B\rbr{D-CA^{-1}B}^{-1} \\ -D^{-1}C\rbr{A-BD^{-1}C}^{-1} & \rbr{D-CA^{-1}B}^{-1} \end{array}\right) \\
\left(\begin{array}{cc} A & B \\ C & D \end{array}\right)^{-1}&=\left(\begin{array}{cc} \rbr{A-BD^{-1}C}^{-1} & -\rbr{A-BD^{-1}C}^{-1}BD^{-1} \\ -\rbr{D-CA^{-1}B}^{-1}CA^{-1} & \rbr{D-CA^{-1}B}^{-1} \end{array}\right)
\end{split}
\end{equation}
then shows, that also the conditions
\begin{align}
V^{(\picount+1)}=&-\frac{2\quadfac^2\timestep}{\hbar}\pdiff{^2\rbr{H^{(\picount+1)}+H^{(\picount+2)}}}{\left.\qkv{\picount+1}\right.^2}-\frac{4\quadfac^4\timestep^2}{\hbar^2}\pdiff{^2H^{(\picount+1)}}{\qkv{\picount+1}\partial\qkv{\picount}}Z^{-1}\pdiff{^2H^{(\picount+1)}}{\qkv{\picount}\partial\qkv{\picount+1}}, \nonumber \\
W^{(\picount+1)}=&\unity{\Sites}-\frac{2\quadfac^2\timestep}{\hbar}\pdiff{^2H^{(\picount+1)}}{\qkv{\picount+1}\partial\pkv{\picount+1}}, \nonumber \\
&-\frac{2\quadfac^2\timestep}{\hbar}\pdiff{^2H^{(\picount+1)}}{\qkv{\picount+1}\partial\qkv{\picount}}Z^{-1}\rbr{\unity{\Sites}+\frac{2\quadfac^2\timestep}{\hbar}\pdiff{^2H^{(\picount+1)}}{\qkv{\picount}\partial\pkv{\picount+1}}}, \\
Y^{(\picount+1)}=&-\frac{2\quadfac^2\timestep}{\hbar}\pdiff{^2H^{(\picount+1)}}{\left.\pkv{\picount+1}\right.^2} \nonumber \\
&-\rbr{\unity{\Sites}+\frac{2\quadfac^2\timestep}{\hbar}\pdiff{^2H^{(\picount+1)}}{\pkv{\picount+1}\partial\qkv{\picount}}}Z^{-1}\rbr{\unity{\Sites}+\frac{2\quadfac^2\timestep}{\hbar}\pdiff{^2H^{(\picount+1)}}{\qkv{\picount}\partial\pkv{\picount+1}}}, \nonumber
\end{align}
hold with $Z=\rbr{V^{(m)}}^{-1}-W^{(\picount)}\rbr{Y^{(\picount)}}^{-1}\left.W^{(\picount)}\right.^{\rm T}$.

Defining $X^{(\picount)}=\rbr{\left.W^{(\picount)}\right.^{\rm T}}^{-1}Y^{(\picount)}\rbr{W^{(\picount)}}^{-1}$ and keeping only terms up to linear order in $\timestep$, these conditions can be written as
\begin{equation}
\begin{split}
V^{(\picount+1)}=&-\frac{2\quadfac^2\timestep}{\hbar}\pdiff{^2\rbr{H^{(\picount+1)}+H^{(\picount+2)}}}{\left.\qkv{\picount+1}\right.^2}, \\
W^{(\picount+1)}=&\unity{\Sites}-\frac{2\quadfac^2\timestep}{\hbar}\pdiff{^2H^{(\picount+1)}}{\qkv{\picount+1}\partial\pkv{\picount+1}}+\frac{2\quadfac^2\timestep}{\hbar}\pdiff{^2H^{(\picount+1)}}{\qkv{\picount+1}\partial\qkv{\picount}}X^{(\picount)}, \\
X^{(\picount+1)}=&X^{(\picount)}-\frac{2\quadfac^2\timestep}{\hbar}\pdiff{^2H^{(\picount+1)}}{{\pkv{\picount+1}}^2}+\frac{2\quadfac^2\timestep}{\hbar}\cbr{\pdiff{}{\pkv{\picount+1}}\rbr{\pdiff{}{\qkv{\picount}}+\pdiff{}{\qkv{\picount+1}}}H^{(\picount+1)}}X^{(\picount)} \\
&+\frac{2\quadfac^2\timestep}{\hbar}X^{(\picount)}\rbr{\pdiff{}{\qkv{\picount}}+\pdiff{}{\qkv{\picount+1}}}\pdiff{H^{(\picount+1)}}{\pkv{\picount+1}} \\
&-\frac{2\quadfac^2\timestep}{\hbar}X^{(\picount)}\cbr{\pdiff{^2\rbr{H^{(\picount+1)}+H^{(\picount+1)}}}{{\qkv{\picount}}^2}+\pdiff{^2H^{(\picount+1)}}{\qkv{\picount}\partial\qkv{\picount+1}}+\pdiff{^2H^{(\picount+1)}}{\qkv{\picount+1}\partial\qkv{\picount}}}X^{(\picount)},
\end{split}
\label{eq:def_X_bosons}
\end{equation}
with the initial condition
\begin{equation}
X^{(1)}=-\frac{2\quadfac^2\timestep}{\hbar}\pdiff{^2H^{(1)}}{{\pkv{1}}^2}.
\end{equation}
In continuous limit $\timestep\to0$, the recursion relation for $X$ gets transformed into a first order differential equation,
\begin{equation}
\dot{X}=-\frac{2\quadfac^2}{\hbar}\pdiff{^2H}{\V{p}^2}+\frac{2\quadfac^2}{\hbar}\pdiff{^2H}{\V{p}\partial\V{q}}X+\frac{2\quadfac^2}{\hbar}X\pdiff{^2H}{\V{q}\partial\V{p}}-\frac{2\quadfac^2}{\hbar}X\pdiff{^2H}{\V{q}^2}X
\end{equation}
with initial condition $X(\ti)=0$ and is solved by
\begin{equation}
X=-\pdiff{\V{q}}{\V{p}(\ti)}\rbr{\pdiff{\V{p}}{\V{p}(\ti)}}^{-1}.
\end{equation}
Therefore, the integral over the fluctuations yields
\begin{align}
&\mathcal{A}_\gamma= \\
&\pilimit\int\frac{\rmd^\Sites y^{(\pisteps)}}{\rbr{2\pi\quadfac}^\Sites}\exp\rbr{\rmi\V{y}^{(\pisteps)}Y^{(\pisteps)}\V{y}^{(\pisteps)}}\prodl{\picount=1}{\pisteps-1}\cbr{\det\rmi\rbr{\begin{array}{cc} -\frac{2\quadfac^2\timestep}{\hbar}\pdiff{^2\rbr{H^{(\picount+1)}+H^{(\picount)}}}{{\qkv{\picount}}^2} & W^{(\picount)} \nonumber \\ \left.W^{(\picount)}\right.^{\rm T} & Y^{(\picount)}\end{array}}}^{-\frac{1}{2}} \nonumber \\
&\qquad=\pilimit\frac{\prodl{\picount=1}{\pisteps-1}\cbr{\det\rbr{\frac{2\quadfac^2\timestep}{\hbar}\pdiff{^2\rbr{H^{(\picount+1)}+H^{(\picount)}}}{{\qkv{\picount}}^2}Y^{(\picount)}+W^{(\picount)}\rbr{Y^{(\picount)}}^{-1}\left.W^{(\picount)}\right.^{\rm T}Y^{(\picount)}}}^{-\frac{1}{2}}}{\sqrt{\det\rbr{-4\rmi\quadfac^2\pi Y^{(\pisteps)}}}} \nonumber \\
%&\qquad=\pilimit\sqrt{\frac{\det\rbr{W^{(\pisteps)}\left.W^{(\pisteps)}\right.^{\rm T}}^{-1}}{\det\rbr{-4\rmi\quadfac^2\pi X^{(\pisteps)}}}}\prodl{\picount=1}{\pisteps-1}\cbr{\det\rbr{\frac{2\quadfac^2\timestep}{\hbar}\pdiff{^2\rbr{H^{(\picount+1)}+H^{(\picount)}}}{{\qkv{\picount}}^2}+\rbr{X^{(\picount)}}^{-1}}\det\rbr{Y^{(\picount)}}}^{-\frac{1}{2}} \nonumber \\
&\qquad=\pilimit\frac{\prodl{\picount=1}{\pisteps-1}\cbr{\det\rbr{\unity{\Sites}+\frac{2\quadfac^2\timestep}{\hbar}\pdiff{^2\rbr{H^{(\picount+1)}+H^{(\picount)}}}{{\qkv{\picount}}^2}X^{(\picount)}}}^{-\frac{1}{2}}}{\sqrt{\det\rbr{-4\rmi\quadfac^2\pi X^{(\pisteps)}}\prodl{\picount=1}{\pisteps}\det\rbr{W^{(\picount)}\left.W^{(\picount)}\right.^{\rm T}}}} \nonumber \\
&\qquad=\pilimit\exp\left\{-\frac{\quadfac^2\timestep}{\hbar}\sul{\picount=1}{\pisteps-1}\Tr\cbr{\pdiff{^2\rbr{H^{(\picount+1)}+H^{(\picount)}}}{{\qkv{\picount}}^2}+2\pdiff{^2H^{(\picount)}}{\qkv{\picount}\partial\qkv{\picount-1}}}X^{(\picount)}\right\} \nonumber \\
&\qquad\qquad\qquad\times\frac{\exp\rbr{\frac{2\quadfac^2\timestep}{\hbar}\sul{\picount=1}{\pisteps}\Tr\pdiff{^2H^{(\picount)}}{\qkv{\picount}\partial\pkv{\picount}}}}{\sqrt{\det\rbr{-4\rmi\quadfac^2\pi X^{(\pisteps)}}}} \nonumber \\
&\qquad=\frac{\exp\rbr{-\frac{\quadfac^2}{\hbar}\intg{\ti}{\tf}{t}\Tr\pdiff{^2\Hcl\rbr{\gVd{\psi}(t),\gV{\psi}(t);t}}{\V{q}(t)^2}X+\frac{\quadfac^2}{\hbar}\intg{\ti}{\tf}{t}\pdiff{^2\Hcl\rbr{\gVd{\psi}(t),\gV{\psi}(t);t}}{\V{q}(t)\partial\V{p}(t)}}}{\sqrt{\det\rbr{-4\rmi\quadfac^2\pi X(\tf)}}}. \nonumber
\end{align}
Using the differential equation for $\partial\V{q}/\partial\V{p}(\ti)$, the exponent can be written as
\begin{align}
\frac{\quadfac^2}{\hbar}&\intg{\ti}{\tf}{t}\Tr\pdiff{^2\Hcl\rbr{\gVd{\psi}(t),\gV{\psi}(t);t}}{\V{q}(t)^2}X=-\frac{\quadfac^2}{\hbar}\intg{\ti}{\tf}{t}\Tr\pdiff{^2\Hcl\rbr{\gVd{\psi}(t),\gV{\psi}(t);t}}{\V{q}(t)^2}\pdiff{\V{q}(t)}{\V{p}(\ti)}\rbr{\pdiff{\V{p}(t)}{\V{p}(\ti)}}^{-1} \nonumber \\
=&\frac{1}{2}\intg{\ti}{\tf}{t}\Tr\rbr{\frac{\rmd}{\rmd t}\pdiff{\V{p}}{\V{p}(\ti)}}\rbr{\pdiff{\V{p}}{\V{p}(\ti)}}^{-1}+\frac{\quadfac^2}{\hbar}\intg{\ti}{\tf}{t}\Tr\pdiff{^2\Hcl\rbr{\gVd{\psi}(t),\gV{\psi}(t);t}}{\V{q}(t)\partial\V{p}(t)} \\
=&\frac{1}{2}\Tr\ln\pdiff{\V{p}(\tf)}{\V{p}(\ti)}+\frac{\quadfac^2}{\hbar}\intg{\ti}{\tf}{t}\Tr\pdiff{^2H}{\V{q}\partial\V{p}}, \nonumber
\end{align}
where in the second steps the equations of motion for $\partial\V{p}(t)/\partial\V{p}(\ti)$ have been used.
%
%I guess, that the final integral is actually always zero \comment{Talk to JD}, such that finally the prefactor becomes simply
Thus, the semiclassical prefactor of the bosonic propagator in quadrature representation is given by the van-Vleck-Gutzwiller determinant
\begin{equation}
\mathcal{A}_{\gamma}=\sqrt{\det\cbr{\frac{1}{-4\rmi\quadfac^2\pi}\rbr{-\pdiff{\V{p}(\ti)}{\qkv{f}}}}}=\sqrt{\det\cbr{\frac{1}{-2\pi\rmi\hbar}\pdiff{^2R_\gamma}{\qkv{f}\partial\qkv{i}}}}.
\end{equation}
\section{The semiclassical prefactor in Fock representation}
\label{app:scl_prefactor_fock}
Starting from \Eref{eq:basis-change_q-n_start} and using Eqns.~(\ref{eq:assymptotics_overlap},\ref{eq:deriv_generating_q-n_q},\ref{eq:deriv_generating_q-n_n},\ref{eq:semiclassical_propagator_quadrature}), the semiclassical propagator in Fock state representation can be written as
\begin{equation}
\begin{split}
\pushleft{K\rbr{\nfv,\niv;\tf,\ti}=} \\
\quad\intg{}{}{^L\qkj{f}{}}\intg{}{}{^L\qkj{i}{}}\sul{\V{s}^{(i)},\V{s}^{(f)}\in\tbr{-1,1}^{\Sites}}{}\cbr{\prodl{\sites=1}{\Sites}\sqrt{\pdiff{^2F\rbr{\qkj{f}{\sites},\nfj{\sites}}}{\qkj{f}{\sites}\partial\nfj{\sites}}\pdiff{^2F\rbr{\qkj{i}{\sites},\nij{\sites}}}{\qkj{i}{\sites}\partial\nij{\sites}}}} \\
\times\sul{\gamma:\qkv{i}\to\qkv{f}}{}\sqrt{\det\frac{1}{(-8\pi^3\rmi\hbar)}\pdiff{^2R_\gamma\rbr{\qkv{f},\qkv{i};\tf,\ti}}{\qkv{f}\partial\qkv{i}}}\rme^{\frac{\rmi}{\hbar}R_{\gamma}\rbr{\qkv{f},\qkv{i};\tf,\ti}} \\
\times\exp\tbr{\rmi\sul{\sites=1}{\Sites}\cbr{s_\sites^{(i)}F\rbr{\qkj{i}{\sites},\nij{\sites}}-s_\sites^{(f)}F\rbr{\qkj{f}{\sites},\nfj{\sites}}}}&.
\end{split}
\label{eq:representation-change_q-n}
\end{equation}
Since the asymptotic formula for the overlap is valid only for the oscillating region $\abs{q}<2\quadfac\sqrt{n+\frac{1}{2}}$, one can substitute $\qkj{i}{1}=2\quadfac\sqrt{\nij{1}+\frac{1}{2}}\cos\thetakj{i}{1}$ and $\qkj{i}{\sites}=2\quadfac\sqrt{\nij{\sites}+\frac{1}{2}}\cos\rbr{\thetakj{i}{\sites}+\thetakj{i}{1}}$ for $\sites=2,\ldots,\Sites$, as well as $\qkj{f}{\sites}=2\quadfac\sqrt{\nfj{\sites}+\frac{1}{2}}\cos\rbr{\thetakj{f}{\sites}+\thetakj{i}{1}}$ for $\sites=1,\ldots,\Sites$. In the following, for reasons that will become obvious after the stationary phase conditions, we will refere to the $\thetakj{i/f}{\sites}$'s as ``phases''. Note that these definitions correspond to those in \Eref{eq:deriv_generating_q-n_n}. Together with the sums over $\V{s}^{(i)}$ and $\V{s}^{(f)}$, the integrals over $\thetakj{i/f}{\sites}$ then run from $-\pi$ to $\pi$ and the phase $\thetakj{i}{1}$ does not enter in the Hamiltonian, which is part of the action.% Moreover, it will turn out to be a global phase, dropping out of the equations of motion. Therefore, the integral over $\thetakj{i}{1}$ can not be evaluated in stationary phase approximation, since it is an integral over a continuous family of trajectories.
%
%However, the remaining $2\Sites-1$ integrals can be evaluated in stationary phase approximation. Here, we will do the two integrations over the final and initial quadratures one after another, starting with the final ones.
%
%The stationary phase condition for the integration over the final phases is given by
%%
%\begin{equation*}
%\pkj{f}{\sites}=2\quadfac\sqrt{\nfj{\sites}+\frac{1}{2}}\sin\rbr{\thetakj{f}{\sites}+\thetakj{i}{1}} \label{eq:spc_q-n_final}
%\end{equation*}
%%
%Thus, the stationary phase approximation selects those classical trajectories $\gV{\psi}(t)$ satisfying the boundary conditions
%%
%\begin{align}
%\abs{\psi_\sites(\tf)}^2=\nfj{\sites}+\frac{1}{2}, \label{eq:bcf_n_app}
%\end{align}
%%
%and therefore justify calling $\thetakj{f}{\sites}$ a ``phase'', since the solutions of the stationary phase conditions are the trajectories with the final phases equal to them.
%
%The evaluation of the exponent at the stationary point will be postponed to when we derived the stationary phase condition for the initial phases.

The second derivative of the exponent in \eref{eq:representation-change_q-n} with respect to the final phases at the stationary point then yields
\begin{equation*}
\begin{split}
&\begin{split}
\pdiff{^2}{\thetakj{f}{\sites}\partial\thetakj{f}{l^\prime}}\Bigg\{\frac{1}{\hbar}R_\gamma\rbr{\qkv{f},\qkv{i};\tf,\ti}+\sul{m=1}{\Sites}\Bigg[\rbr{\nfj{m}+\frac{1}{2}}\rbr{\thetakj{f}{m}+\thetakj{i}{1}} \\
-\sqrt{\nfj{m}+\frac{1}{2}}\cos\rbr{\thetakj{f}{m}+\thetakj{i}{1}}\sin\rbr{\thetakj{f}{m}+\thetakj{i}{1}}\Bigg]\Bigg\}
\end{split} \\
&\quad=\sqrt{\nfj{\sites}+\frac{1}{2}}\sin\rbr{\thetakj{f}{\sites}+\thetakj{i}{1}}\cbr{2\delta_{\sites\sites^\prime}\sqrt{\nfj{\sites}+\frac{1}{2}}\cos\rbr{\thetakj{f}{\sites}+\thetakj{i}{1}}-\frac{1}{\quadfac}\pdiff{\pkj{f}{\sites}}{\thetakj{f}{\sites^\prime}}} \\
&\quad=-\pdiff{}{\thetakj{f}{\sites^\prime}}\frac{1}{4\quadfac^2}\cbr{\rbr{\qkj{f}{\sites}}^2+\rbr{\pkj{f}{\sites}}^2}=-\rbr{\pdiff{\thetakj{f}{\sites^\prime}}{\nfj{\sites}}}^{-1}.
\end{split}
\end{equation*}
Therefore, together with the prefactor given by the second derivative of the generating function, the semiclassical prefactor after performing the integration over the final phases is given by
\begin{equation}
\sqrt{\prodl{\sites=1}{\Sites}\pdiff{\qkj{i}{\sites}}{\theta_{\sites}^{(i)}}}\sqrt{\frac{1}{\rbr{-4\pi^2\rmi\hbar}^\Sites}\det\pdiff{^2R_\gamma\rbr{\nfv,\qkv{i};\tf,\ti}}{\nfv\partial\qkv{i}}},
\label{eq:prefactor_mixed_n-q}
\end{equation}
where we defined
\begin{equation*}
\begin{split}
R_\gamma\rbr{\nfv,\qkv{i};\tf,\ti}=R_\gamma\rbr{\qkv{f},\qkv{i};\tf,\ti}+\hbar\sul{\sites=1}{\Sites}\Bigg[\rbr{\nfj{\sites}+\frac{1}{2}}\rbr{\thetakj{f}{\sites}+\thetakj{i}{1}} \\
-\sqrt{\nfj{\sites}+\frac{1}{2}}\cos\rbr{\thetakj{f}{\sites}+\thetakj{i}{1}}\sin\rbr{\thetakj{f}{\sites}+\thetakj{i}{1}}&\Bigg]
\end{split}
\end{equation*}
evaluated at the stationary point.

Noticing that
\begin{equation*}
\pdiff{R_\gamma\rbr{\nfv,\qkv{i};\tf,\ti}}{\nfv}=\hbar\rbr{\thetakv{f}+\thetakj{i}{1}\V{1}},
\end{equation*}
where $\V{1}$ is the vector, for which every entry is equal to one, one can write \eref{eq:prefactor_mixed_n-q} also as
\begin{equation}
\sqrt{\det\frac{1}{\rbr{-4\pi^2\rmi}}\pdiff{\rbr{\gV{\theta}^{(f)}+\theta_1^{(i)}\V{1}}}{\gV{\theta}^{(i)}}}.
\end{equation}
%
%Now, the stationary phase condition for the integration over the initial phases, save for $\theta_1^{(i)}$, yields
%\begin{equation}
%\pkj{i}{\sites}=2\quadfac\sqrt{\nij{\sites}+\frac{1}{2}}\sin\rbr{\thetakj{i}{\sites}+\thetakj{i}{1}},
%\end{equation}
%%
%which corresponds to
%%
%\begin{equation}
%\abs{\psi_\sites(\ti)}^2=\nij{\sites}+\frac{1}{2}. \label{eq:bci_n_app}
%\end{equation}
%%
%Note that with \eref{eq:bcf_n_app} and \eref{eq:bci_n_app} the conservation of the total number of particles together with the fact that $\qkj{i}{1}=\sqrt{\nij{1}+1/2}\cos\thetakj{i}{1}$ already requires that
%%
%\begin{equation}
%\psi_1(\ti)=\sqrt{\nij{1}+\frac{1}{2}}\exp\rbr{\rmi\thetakj{i}{1}}.
%\label{eq:bci_n_firstcomponent}
%\end{equation}
%%
%Moreover, we see that $\thetakj{i}{1}$ indeed is a global phase, which does neither enter the equations of motion, nor the Hamiltonian. Therefore, the solutions to the equations of motion under the boundary conditions \eref{eq:bcf_n_app}, \eref{eq:bci_n_app} and \eref{eq:bci_n_firstcomponent} are the trajectories with the initial phase of the first component fixed to a certain $\varphi$, \eg $\varphi=0$, and multiplied by a global phase factor $\exp\rbr{\rmi\thetakj{i}{1}-\rmi\varphi}$.
%
%With this, one can now evaluate the exponent at the stationary points, which will then give the classical action in Fock representation.
Substituting $q_\sites(t)=2\sqrt{n_\sites(t)+\frac{1}{2}}\cos\rbr{\theta_\sites(t)+\thetakj{i}{1}}$ and $p_\sites(t)=2\quadfac\sqrt{n_\sites(t)+\frac{1}{2}}\sin\rbr{\theta_\sites(t)+\thetakj{i}{1}}$ in the kinetic part of the action, as required by the stationary phase condition \Eref{eq:bc_fock}, in consideration of the preservation of the total number of particles by the classical equations of motion yields for the classical action in Fock state representation
\begin{align}
R_\gamma\rbr{\nfv,\niv;\tf,\ti}=\intg{\ti}{\tf}{t}\cbr{\hbar\gV{\theta}(t)\cdot\dot{\V{n}}(t)-\Hcl\rbr{\gVd{\psi}(t),\gV{\psi}(t);t}}.
\end{align}
Moreover we have,
\begin{equation}
\begin{split}
\pdiff{^2}{\thetakj{i}{\sites}\partial\thetakj{i}{\sites^\prime}}\Bigg\{\frac{1}{\hbar}R\rbr{\nfv,\qkv{i};\tf,\ti}-\sul{m=1}{\Sites}\Bigg[\rbr{\nij{m}+\frac{1}{2}}\rbr{\thetakj{i}{m}+\thetakj{i}{1}} \\
-\sqrt{\nij{m}+\frac{1}{2}}\cos\rbr{\thetakj{i}{m}+\thetakj{i}{1}}\sin\rbr{\thetakj{i}{m}+\thetakj{i}{1}}\Bigg]\Bigg\}&=\rbr{\pdiff{\thetakj{i}{\sites^\prime}}{\nij{\sites}}}^{-1}.
\end{split}
\end{equation}
and therefore
\begin{equation}
\begin{split}
\pushleft{K\rbr{\nfv,\niv;\tf,\ti}=} \\
\qquad\frac{1}{2\pi}\intg{0}{2\pi}{\thetakj{i}{1}}\sul{\gamma:\nfv\to\niv}{}\sqrt{\frac{1}{-2\pi\rmi}\det\rbr{\pdiff{\thetakj{i}{\sites}}{\nij{\sites^\prime}}}_{\sites,\sites^\prime=2,\ldots,\Sites}\det\pdiff{\rbr{\thetakv{f}+\thetakj{i}{1}\V{1}}}{\thetakv{i}}} \\
\times\exp\cbr{\frac{\rmi}{\hbar}R_\gamma\rbr{\nfv,\niv;\tf,\ti}}&.
\end{split}
\label{eq:propagator_fock_almost}
\end{equation}
It is important to notice that the two matrices in the prefactor have different dimensions. The first one is a $(\Sites-1)\times(\Sites-1)$-matrix, while the second one is $\Sites\times\Sites$ dimensional. However, since $\thetakv{f}$ does not depend on $\thetakj{i}{1}$, the second one has the form
\begin{equation}
\left(\V{1}\quad\pdiff{\thetakv{f}}{\tilde{\gV{\theta}}^{(i)}},\right)
\end{equation}
where $\tilde{\gV{\theta}}^{(i)}$ is the vector resulting from $\thetakv{i}$ by skipping its first entry. Moreover, we can match the dimensions of the two matrices by using
\begin{equation}
\det\rbr{\pdiff{\tilde{\gV{\theta}}^{(i)}}{\tilde{\V{n}}^{(i)}}}=\det\rbr{
\begin{array}{cc}
1 & 0 \\
0 & \pdiff{\tilde{\gV{\theta}}^{(i)}}{\tilde{\V{n}}^{(i)}}
\end{array}}
\end{equation}
Then,
\begin{equation}
\left(\V{1}\quad\pdiff{\thetakv{f}}{\tilde{\gV{\theta}}^{(i)}},\right)\rbr{
\begin{array}{cc}
1 & 0 \\
0 & \pdiff{\tilde{\gV{\theta}}^{(i)}}{\tilde{\V{n}}^{(i)}}
\end{array}}=\rbr{\V{1}\quad\pdiff{\thetakv{f}}{\tilde{\V{n}}^{(i)}}}.
\end{equation}
The determinant of this matrix however can be further simplified by using Sylvester's determinant theorem, which states that for an $n\times m$ matrix $A$ and an $m\times n$ matrix $B$
\begin{equation}
\det\rbr{\unity{n}+AB}=\det\rbr{\unity{m}+BA},
\end{equation}
where $\unity{n}$ is the $n\times n$ unit matrix and applying it to the right hand side of
\begin{equation}
\det\rbr{\begin{array}{cc}
1 & \pdiff{\thetakj{f}{1}}{\tilde{\V{n}}^{(i)}} \\
\V{1} & \pdiff{\tilde{\gV{\theta}}^{(f)}}{\tilde{\V{n}}^{(i)}}
\end{array}}=
\det\rbr{
\begin{array}{cc}
1 & \pdiff{\thetakj{f}{1}}{\tilde{\V{n}}^{(i)}} \\
\V{0} & \pdiff{\tilde{\gV{\theta}}^{(f)}}{\tilde{\V{n}}^{(i)}}
\end{array}}\cbr{\unity{L}+\rbr{\begin{array}{cc}
-\pdiff{\thetakj{f}{1}}{\tilde{\V{n}}^{(i)}}\rbr{\pdiff{\tilde{\gV{\theta}}^{(f)}}{\tilde{\V{n}}^{(i)}}}^{-1}\V{1} & \V{0}^{\mathrm{T}} \\
\rbr{\pdiff{\tilde{\gV{\theta}}^{(f)}}{\tilde{\V{n}}^{(i)}}}^{-1}\V{1} & 0
\end{array}}}
\end{equation}
With this, one can show that
\begin{equation}
\det\rbr{\V{1}\quad\pdiff{\gV{\theta}^{(f)}}{\tilde{\V{n}}^{(i)}}}=\det\pdiff{\tilde{\gV{\theta}}^{(f)}}{\tilde{\V{n}}^{(i)}}=\det\hbar\pdiff{^2R_{\gamma}\rbr{\nfv,\niv;\tf,\ti}}{\tilde{\V{n}}^{(f)}\tilde{\V{n}}^{(i)}}.
\end{equation}
Finally, we see that the integrand in \eref{eq:propagator_fock_almost} is independent of $\theta_1^{(i)}$, such that the final result reads
\begin{equation}
\begin{split}
K\rbr{\nfv,\niv;\tf,\ti}=\sul{\gamma:\niv\to\nfv}{}\sqrt{{\det}^{\prime}\frac{1}{\rbr{-2\pi\rmi\hbar}}\pdiff{^2R_\gamma\rbr{\nfv,\niv;\tf,\ti}}{\V{n}^{(f)}\partial\V{n}^{(i)}}} \\
\times\exp\tbr{\frac{\rmi}{\hbar}R_\gamma\rbr{\nfv,\niv;\tf,\ti}}&.
\end{split}
\end{equation}
\end{appendix}
\bibliography{tom}
\bibliographystyle{rsta}
\end{document}